# Large Area Electrically Tunable Lenses Based on Metasurfaces and Dielectric Elastomer Actuators

**Alan She[1], Shuyan Zhang[1], Samuel Shian[1], David R. Clarke[1,*], Federico Capasso[1,*]**
**[1]John A. Paulson School of Engineering and Applied Sciences, Harvard University, 9 Oxford Street, Cambridge, MA 02138, USA**
***e-mail: clarke@seas.harvard.edu; capasso@seas.harvard.edu**

**Tunable optical devices, in particular, varifocal lenses, have important applications in various fields, including imaging and adaptive vision. Recent advances in metasurfaces, which control the wavefront of light using subwavelength-spaced nanostructures, open up new opportunities to replace bulk optical devices, with thin, flat, lightweight devices. We have demonstrated for the first time an electrically tunable flat lens, based on large area metasurfaces combined with a dielectric elastomer actuator having inline transparent electrodes, that is capable of simultaneously performing focal length tuning (>100%) as well as dynamic corrections, including astigmatism and image shift. This offers control versatility to flat optics hitherto only possible in electron microscopes. The water-based transfer process, which we describe, also enables wide compatibility across materials. The combination of metasurface optics and dielectric elastomer actuators enables a new, versatile platform for creating all kinds of tunable optical devices, through the design of tunable phase, amplitude, and polarization profiles.**

There has been a wide variety of work on tunable optical devices, in particular, tunable focus lenses (i.e., varifocal lenses), with important applications in imaging and adaptive vision. State-of-the-art technologies include the following: Conventional optical zoom, involving translating rigid elements, is bulky and inertially limited in speed[1,2]. Liquid crystal spatial light modulators with transparent electrodes made of indium tin oxide are high speed and use low operating voltage but are limited in resolution and polarization dependence[3,4]. Fluid based tunable lenses are high speed and have a wide tuning range, but the control of the exact surface curvature is difficult, and coma has been observed when placing the lens vertically, due to gravity[5-10]. Tunable acoustic gradient index lenses use acoustic waves to generate an oscillating refractive index profile, which can be tuned at high speeds (~μs) but must operate in a stroboscopic mode[11,12].

Recent advances in metasurfaces open up new directions to replace bulk optical devices, with thin, flat, lightweight devices. Metasurfaces control the wavefront of light by using subwavelength-spaced arrays of fixed optical phase shifters, amplitude modulators, or polarization changing elements, patterned on a surface to introduce a desired spatial distribution of optical phase, amplitude, or polarization. By engineering the properties of each metasurface element, one can shape the wavefront of the transmitted, reflected, or scattered light[13]. Based on this concept, various functionalities have already been demonstrated including lenses, blazed gratings, axicons, vortex plates, wave plates, and polarimeters[14-16]. However, the majority of proposed metasurface devices are static, limiting their potential applications.

Mechanically-tunable metasurfaces have been reported recently, in which metasurfaces are embedded in stretchable substrates, such as a 1.7X zoom lens[17], a reconfigurable hologram[18], and tunable lens with focusing efficiency above 50%[19]. However, all of these were much less than a millimeter in size and required external apparatuses to generate mechanical strain, restricting their applications.

In this paper, we introduce a new approach, in which metasurfaces can be made electrically tunable by combining them with a class of electroactive polymers known as dielectric elastomer actuators (DEA)[8]. Large biaxial and reversible strains at relatively high speed can be produced by applying an electric field and the biaxial strain is directly compatible with 2D metasurfaces. By using different electrode configurations, they can be used to alter the size, shape, and lateral translations of a metasurface bonded to the elastomer. We show the potential of this approach in performing simultaneous tuning of focal length, astigmatism correction, and image shift control. This brings the familiar tuning capabilities of electron microscopes to flat optics. It also enables long sought-after applications, in which dynamic and high-speed tuning can be achieved with voltage-resolved precision in an analog or digital manner at the millisecond time scale.

## Theory

A flat lens can be constructed using the following hyperboloidal phase profile:

$$\phi(r) = \pm k\left(\sqrt{r^2+f^2}-f\right), \quad (1)$$

where $k$ is the wavenumber ($k = 2\pi n/\lambda$, where $n$ is the refractive index of the medium in which the lens is immersed and $\lambda$ is wavelength), $r$ is radial position, $f$ is focal length, and the sign is positive or negative for diverging or converging lenses, respectively (Fig. 1a). This phase profile focuses light free of spherical aberrations for normal incidence (infinity-corrected) illumination. We combine the metasurface lens with a DEA, to make the metasurface tunable, which was, by itself, a static device.

The DEA, sometimes referred to as “artificial muscles” in soft robotics, is effectively a compliant parallel plate capacitor[20]. By using a soft elastomer (e.g., polyacrylate and silicone rubbers), as a dielectric together with transparent, stretchable electrodes, the dielectric is compressed in the electric field direction (thickness direction) when a voltage is applied. As elastomers conserve volume on deformation (i.e., Poisson ratio close to 0.5), the thinning results in lateral or areal expansion that can be very large, up to 500%[21]. The attainable actuation strain is a function of the electrostatically-induced Maxwell stress, the constitutive deformation behavior of the elastomer, and mechanical configuration of the elastomers and electrodes[22].

By bonding the metasurface lens to the dielectric elastomer, the in-plane (lateral) deformation of the metasurface lens is coupled to the voltage-induced stretching of the elastomer, which is uniform over the electrode area. This uniform stretching effectively scales the lateral coordinates of the lens by a stretch factor, $s$ ($s = 1 + \varepsilon_L$, where $\varepsilon_L$ is lateral strain), such that $r \rightarrow r/s$:

$$\phi_{stretched}(r,s) = \pm k\left(\sqrt{(r/s)^2+{f_0}^2}-f_0\right), \quad (2)$$

where $f_0$ is the focal length without actuation, provided that metasurface elements retain their original, individual phase responses, and only their separation is increased (Fig. 1b). While this stretched phase profile is not of the same form as Eq. 1 (a radially varying stretch profile, $s_{radial}(r)$, would be required, see Supplementary Section 1), it is closely approximated (Supplementary Fig. S1a,b) by the following ideal phase profile:

$$\phi_{ideal}(r,s) = \pm k\left(\sqrt{r^2+(s^2 f_0)}-s^2 f_0\right), \quad (3)$$

in which the focal length of Eq. 1 is replaced by stretch-dependent focal length, $f(s) = s^2 f_0$. $\phi_{ideal}$ produces a focal length change that scales quadratically with stretch while maintaining the hyperboloidal shape necessary to be free of spherical aberration. The voltage applied to a DEA alters the focal length in the following way:

$$\frac{f}{f_0} = \frac{1}{1-(\varepsilon/Y)(V/t)^2}\ , \tag{4}$$

where $V$ is and voltage, and $\varepsilon$, $Y$, and $t$ are the permittivity, Young's modulus, and dielectric layer thickness, respectively (Supplementary Section 2).

In tunable systems, optical aberrations may be augmented or reduced by tuning. With metasurface lenses, spherical aberration using normal incidence illumination is corrected by design. However, when such a metasurface is deformed, the resulting aberration is not obvious. The wavefront aberration function ($WAF = \phi_{stretched} - \phi_{ideal}$) quantifies the deviation of the resulting phase profile of the stretched metasurface from the ideal phase profile:

$$\begin{aligned} WAF &= \pm k\left(\sqrt{(r/s)^2 + f_0^{\ 2}} - f_0^{\ 2}\right) \mp k\left(\sqrt{r^2 + (s^2 f_0)^2} - s^2 f_0\right) \\ &= \pm k\frac{(1-s^2)r^4}{8f^3 s^6} + O(r^6)\,. \end{aligned} \tag{5}$$

This equation shows that as the uniform stretch is increased from $s = 1$, the WAF increases from 0 until reaching a maximum aberration (Supplementary Fig. S1c) at $s \approx 1.22$ (e.g., for a lens ⌀6 mm, f = 50 mm, the maximum aberration is <0.05 rad at the edge). Upon further stretching, a built-in suppression of spherical aberration comes into effect, in which Eq. 5 decays following a quartic function, $(r/s)^4$. This allows for highly tunable lens devices with excellent immunity to aberration.

In practice, an infinite number of aberrations exist and can be quantified in terms of Zernike polynomials (Supplementary Table S1). In most cases, the lowest eight terms are sufficient: piston, tip, tilt, defocus, oblique astigmatism, vertical astigmatism, vertical coma, and horizontal coma. Because Zernike terms are linear and orthogonal, specific or multiple optical aberrations can be targeted and tuned by introducing the appropriate strain field to the phase profile. The application of stress of a particular configuration induces a strain field, resulting in a displacement field. The displacement field can be regarded as the sum of deformation ($\overline{A}$) and a rigid body displacement ($\overline{B}$) components, in which the transformed coordinates can be expressed in the following way: $\bar{x}' = \overline{A}(\bar{x}, \eta) + \overline{B}(t)$, where $\eta$ is a parameterization (e.g., stress, voltage) and $\bar{x}$ and $\bar{x}'$ are the original and transformed coordinates, respectively[23]. For metasurfaces, this reduces to a 2-D problem, in which the phase profile coordinates transform: $\phi(x, y) \rightarrow \phi(x', y')$. The rigid body displacement is simply a shift in the x,y plane of the entire phase profile (Fig. 1d), while the deformation changes the shape or size, opening up opportunities for generating new phase profiles through tuning the strain field. In the case of asymmetric biaxial strain, the lateral coordinates of the phase profile transform as follows $(x, y) \rightarrow (x/s_x, y/s_y)$:

$$\phi = \pm k\left(\sqrt{(x/s_x)^2 + (y/s_y)^2 + f_0^{\ 2}} - f_0\right), \tag{6}$$

where $s_x \neq s_y$ ($s_x = s_y$ is uniform stretching). Here light propagating along two perpendicular planes experience different focal lengths, which is astigmatism (Fig. 1c).

Using the voltage-induced strain field, it is possible to create the optical analog of image shift and stigmators found in scanning and transmission electron microscopes, which shape the electron beam dynamically[24]. In optical systems, astigmatism can be performance-limiting, and sources of astigmatism often stem from geometric imperfections, such as misaligned, malformed, or strained lenses. Optical shift and astigmatism tuning is particularly useful in systems with that components move or deliberately utilize these parameters to extract additional information.

## Device design

We designed and fabricated a polarization-insensitive, converging metasurface lens with a diameter of 20 mm and a focal length of 50 mm for a wavelength of 1550 nm based on Eq. 1. The metasurface consisted of amorphous (a)-Si circular posts (Fig. 2b and Methods). The data file for the photomask pattern generator was created using an algorithm (Supplementary Fig. S2c) we recently developed for efficiently generating large area metasurface designs with an extremely large number of densely-packed polygons[25]. We constructed the DEA using transparent polyacrylate elastomers (VHB 4905, 3M) with stretchable-transparent electrodes made of single-walled carbon nanotubes (SWCNT)[26]. These were measured to exhibit large transparency windows in the visible, near-infrared, and mid-infrared spectra (Supplementary Fig. S3).

The lens was fabricated using photolithography in a prepared film stack (see Fig. 2a and Methods). By attaching the DEA to the sample surface and dissolving the sacrificial layer in water, the metasurface was transferred to the DEA. Electrical leads were then attached to apply control voltages. The final device (Fig. 3a) includes a metasurface lens with a diameter of 6 mm.

Focal length tuning was implemented by applying a voltage through the center electrode $V_5$ for increasing focal length or from $V_1$ to $V_4$ together for decreasing focal length (Fig. 1b and 3b), corresponding to lateral expansion or contraction of the post spacings, respectively. The control of vertical astigmatism in the x, y directions ("x,y-stigmators") was implemented by activating opposing pairs of electrodes (Fig. 1c and 3b). Image shift was implemented by activating one peripheral electrode: $V_1$ through $V_4$, such that its expansion causes the entire metasurface to shift in space (Fig. 1d and 3b). Any combination of actuations, each implemented using different voltages, is also possible, depending on the desired control over the strain field to generate certain optical results. For example, the defocus, stigmators, and shifters can be utilized simultaneously to exert multi-parameter control.

## Characterization

Two types of devices were fabricated and measured, a single layer (SL) and double layer (DL) device (Fig. 3a). While the SL demonstrated better tunability than the DL, most of the data presented are of the DL, due to it being higher quality.

DEA induced stretch was observed from optical microscope images of the center of the lens. The Fourier transforms (FT) of the DL images (Fig. 4a) were used to quantify the stretch (Fig. 4b), in which we measured the radii of the first order annulus of FT, corresponding to the reciprocal space representation of the radially symmetric post-to-post separation. With the same applied voltage, the DL was stretched less than SL, on account of the greater stiffness introduced by the larger intermediate elastomer layer. The maximum voltage used was 3 kV, which produced s = 1.41 (SL) and 1.15 (DL).

The focal lengths were measured by scanning a camera along the z-axis for different applied voltages (see Methods). From f=50 mm without actuation, the DL focal length was tuned by 15 mm (30%, $\Delta f/f_0$) as the voltage was increased from 0 to 3 kV (Fig. 4b) and closely followed the predicted relation with voltage (Eq. 4 and Supplementary Fig. S6). The SL displayed a greater, 107% focal length modulation upon varying the voltage from 0 to 3 kV (Fig. 4b, inset). The focusing efficiency, defined as the ratio of the focused optical power and the incident power (Supplementary Fig. S4c), was measured for the DL and showed a high average efficiency of 62.5% with minimal variation throughout the tuning range (Fig. 4c). For comparison, the measured focusing efficiency of the metasurface before transfer (fabricated on a fused silica wafer) was 91%, and the difference was attributed scattering and absorption by the DEA as well as imperfections in the transfer process. Images of the focus were taken to investigate the effect of aberrations upon tuning (Fig. 4d). Since the sizes of the beam (⌀7 mm) and lens (⌀6 mm) were comparable, the illumination profile impinging on the lens was Gaussian rather than a plane wave, causing the Airy pattern to not manifest by apodization. In the absence of any applied voltage, the measured focal spot size ($1/e^2$ full beam waist) was 34.4±1.1 μm, compared to the theoretical diffraction-limited spot size of 21.4 μm (⌀6 mm, f=50 mm, $M^2$=1.3). At 1.0 kV, the measured focal spot size was 37.7±2.8 μm, compared to a diffraction-limited spot size of 22.7 μm (⌀6 mm, f=53 mm, $M^2$=1.3). Possible reasons for this difference include errors introduced during fabrication and the transfer process. In particular, it is likely that a non-uniform strain field was created as the elastomer membrane was initially pressed onto the metasurface, resulting in small distortions in the final device. Optical performance can be improved by optimizing these procedures. The focal spot sizes were within their measurement errors indicating that the focal spot size, although not diffraction-limited, was maintained on tuning. Due to setup constraints (Supplementary Fig. S4b), images of the focal spots at higher voltages (> 1 kV) were obtained directly by the camera without magnification (Fig. 4b, bottom). It was clearly seen that the focal spot maintains its shape even at high voltages.

We measured the x stigmator (Fig. 4a and 4e), which squeezed the phase profile into an elliptical shape by contraction and expansion in the x and y directions, respectively. The image FT (Fig. 4a) was then used to quantitatively measure the strain field, showing asymmetric yet spatially uniform biaxial strain: a requirement for good stigmator performance. The resulting astigmatism tuning was calculated by taking the Zernike transform of stretched phase profile (Fig. 4e).

We measured the x,y-shift and were able to shift the image in the up, down, left and right directions by applying appropriate control voltages (Fig. 4f). The observed asymmetry of the control is most likely due to some residual asymmetrical stiffness around the periphery of the device. This asymmetry occurred during fabrication, as three different structures needed to be concentrically aligned: the outer frame, electrodes, and metasurface. Precise alignment between these structures would improve the symmetry of control. Shift-induced distortion of the metasurface was minimal (Supplementary Fig. S5).

The reliability of the device was tested with a sinusoidal voltage from 2 to 100 Hz at an amplitude of 2.5 kV to continuously vary the focal length (Supplementary Movie S1). More than 1000 cycles without failure were achieved. By applying a voltage square wave, we were able to measure the response time of our devices to be 33±3 ms (Supplementary Fig. S7). The response time is mainly limited by the viscoelasticity of the elastomer, the charge transfer and dissipation time of the SWCNT electrodes. It is expected that both can be improved by using better elastomer materials and reducing the electrical resistance of the electrodes. Dielectric breakdown of the device was also investigated. The lens can be tuned by increasing the voltage until breakdown occurs, upon which current begins to flow through the dielectric, damaging the device. Our devices broke down at around 3.5 kV. Interestingly, the electrical breakdown was a “soft” breakdown associated with local burning through the elastomer and the same devices were able to resume normal operation after disconnecting the power and allowing the device to rest for several hours. This self-healing feature is attributed to the burning and subsequent clearing of SWCNT electrodes around the breakdown location, preventing further electrical shorting.

The current device can be improved in the following ways: Reducing the thickness of the elastomer could reduce the voltage requirement[27] (possibly to less than 12 V). Miniature high voltage components are commonly found in devices, such as cell phones, including flash modules, piezo actuators, and surface acoustic wave transducers. Another consideration is to decrease the response time to potentially μs time scale[7,28]. Also, optical performance may be improved by optimizing the transfer process as well as using other DEA materials[29].

## Conclusion

We have demonstrated for the first time an electrically-tunable metasurface lens whose focal length, astigmatism and shift can be simultaneously controlled by applied voltage signals to a mechanically coupled, transparent dielectric elastomer actuator. The focal length scales as the quadratic function of stretch ($\propto s^2$), yielding large changes in focal length for small strains. It brings into focus embedded optical zoom and optical image stabilization for chip-scale image sensors (e.g. cell phone cameras) as well as optical zoom and adaptive focus with lightweight form factors for head mounted optics, such as everyday eyeglasses, virtual reality and augmented reality hardware. In other applications, it allows for optical zoom and focal plane scanning for cameras, telescopes, and microscopes without the need for motorized parts. Furthermore, its flat construction and inherently lateral actuation allow for highly stackable systems, such as compound optics. While the device presented used rigid metasurface elements, non-rigid metasurface elements individually tunable with strain can also be designed. The combination of metasurface optics and DEAs enables a versatile platform for tunable optical devices, including tunable phase, amplitude, and polarization profiles, through the electrical control of the strain field in the optical layer, making it possible to bring tunability long familiar in electron optics to flat optics.

## Acknowledgements

This work was financially supported by the Air Force Office of Scientific Research (AFOSR) under MURI: FA9550-12-1-0389. A.S. thanks the Charles Stark Draper Laboratory for support through the Draper Laboratory Fellowship. S.Z. thanks the A*STAR Singapore for support through the National Science Scholarship. The work of S.S. was supported in part by the National Science Foundation through the grant CMMI-1333835 and by the MRSEC program of the National Science Foundation under award number DMR 1420570. F.C. gratefully acknowledges a gift from Huawei Inc. under its HIRP FLAGSHIP program. This work was performed in part at the Center for Nanoscale Systems (CNS), a member of the National Nanotechnology Infrastructure Network (NNIN), which is supported by the National Science Foundation under NSF award no. ECS-0335765. CNS is part of Harvard University. This work was performed in part at the Cornell NanoScale Facility (CNF), a member of the National Nanotechnology Coordinated Infrastructure (NNCI), which is supported by the National Science Foundation (Grant ECCS-1542081). The authors would also like to express their gratitude to R. M. Diebold for helpful discussions about dielectric elastomers; G. Zhong for generous help in using CNS facilities; and J. Treichler, A. Windsor, G. Bordonaro, K. Musa, and D. Botsch for their generous help in using CNF facilities.

## Author contributions

A.S., S.Z., and S.S. fabricated the materials and devices; A.S. and S.Z. performed the simulations, measurements and analysis with assistance from S.S.; A.S. and S.Z. wrote the manuscript. F.C. and D.R.C. supervised the project; All authors discussed the results and commented on the manuscript.

## Competing financial interests

The authors declare no competing financial interests.

## Materials & Correspondence

Correspondence and requests for materials should be addressed to F.C. and D.R.C.

## Methods

**Metasurface lens design.** The unit cell of the design is shown in the Supplementary Fig. S2a. The height of the posts is h = 950 nm. By varying the diameter of the posts (d = 810 - 990 nm), a phase coverage of close to $2\pi$ and a high, uniform transmission amplitude response are achieved (see Supplementary Fig. S2b). These data were used as a lookup table to digitize the phase profile. Post diameters with low transmission values (e.g., d = 860 and 870 nm) were excluded from the lookup table. Given the circular shape of the post structures, the phase and amplitude responses are independent of the polarization of the incoming light. The phase profile was realized by subwavelength antennas with fixed edge-to-edge separation (see Supplementary Fig. S2c), by which the placement of antennas is made denser than that with the conventional fixed center-to-center separation. Hence the size of the unit cell ($u$) is equal to the sum of the post diameter ($d$) and the constant edge-to-edge spacing ($e$): $u = d + e$. In our design, we chose the edge-to-edge spacing to be 650 nm, which was determined by the feature size of the stepper we used as well as the length scale to avoid interaction between neighboring antennas.

**Selection of materials.** Although VHB (VHB 4905, 3M) is not an optimized DEA material, we chose to use it as the elastomer because it is transparent, sticky, easily available, and is capable of large strain[29]. Silicone-based elastomers, which are also transparent, offer more precise control, larger temperature stability and very low hysteresis, but their use requires additional processing steps.

In order to transfer a metasurface from its substrate to the elastomer membrane, a sacrificial layer was used. It is important for the sacrificial layer to be soluble in a solvent orthogonal to both the metasurface and the elastomer, such that the solvent only dissolves the sacrificial layer but not the metasurface nor elastomer. We chose germanium dioxide ($GeO_2$) as the sacrificial layer, since it can be readily dissolved in water. Water enables versatility of the process to be used with a wide variety of metasurface materials and membrane materials, as well as in other applications.

**Fabrication.** A film stack was first prepared for nanofabrication. Starting with a silicon wafer, a 0.4 μm layer of elemental germanium (Ge) was deposited by electron beam evaporation. The sample was then placed in a furnace (Tystar Tytan) for dry oxidization in the presence of $O_2$ at atmospheric pressures at 550 °C for 3 hours, which converted the layer of Ge into $GeO_2$, completing the sacrificial layer. We found that the $GeO_2$ layer made by dry oxidizing Ge was significantly more soluble in water than layers deposited by either electron evaporation and thermal evaporation directly using $GeO_2$ as the source material, possibly due to formation of $GeO_x$ species at high temperatures. $GeO_2$ also dissolves much faster in water than Ge in hydrogen peroxide ($H_2O_2$) and water solution, in which the rate-limiting step is the oxidation of Ge into $GeO_2$ by $H_2O_2$. The metasurface itself is composed of nano-posts made of amorphous silicon (a-Si) with height of 950 nm, so a layer of a-Si with the corresponding thickness was deposited using plasma-enhanced chemical vapor deposition (PECVD).

In preparation for photolithography, the sample surface was first spin coated with the adhesion promoter, hexamethyldisilazane (HMDS), at 4000 rpm. Next, a 1 um layer of i-line photoresist (SPR700-1.0, DOW) was spun-coated and soft baked at 95 °C for 60 s. Over the photoresist, a layer of photo-bleachable contrast enhancement material (CEM365iS, ShinEtsuMicroSi) was spun-coated at 4000 rpm to improve feature contrast for the following stepper exposure.

The metasurface design (Fig. S2) was patterned into the photoresist using stepper photolithography, which allowed us to produce large area metasurface lenses with high yield. A quartz photomask was patterned using a high resolution laser lithography system (Heidelberg DWL2000). This photomask was then used as the reticle in a 5x reduction i-line stepper (GCA AS200 AutoStep) to expose the photoresist. After exposure, the CEM365iS layer was removed by spraying deionized water (DIW) and spin drying. The sample was then post exposure baked (PEB) at 115 °C for 60 s. The photoresist was then soaked in developer (MF CD26) for 90 s and rinsed in DIW. A mild $O_2$ plasma descum was performed to improve pattern fidelity.

The photoresist pattern was used as the etch mask to create the metasurface in the a-Si layer. The sample was etched using an inductively coupled plasma (ICP) reactive ion etch (RIE) system (STS MPX/LPX ICP RIE), in which the etchant gases used were perfluorocyclobutane ($C_4F_8$) and sulfur hexafluoride ($SF_6$). The pattern was etched completely through the a-Si layer and as much as 100 nm into the sacrificial layer.

Finally, the photoresist was removed by soaking the sample in N-methyl-2-pyrrolidone (NMP) solution (Remover PG, Microchem) for 8 hours followed by a dry resist strip in a high temperature (200 °C) and high power (500 W) $O_2$ plasma asher (Matrix Plasma Asher) for 20 s, leaving a-Si and $GeO_2$ exposed.

To create the DEA, a membrane of elastomer, VHB, was uniformly biaxially stretched (4x, linear) and mounted by its own adhesion on a rigid circular plastic ring. SWCNT electrodes were then applied to either side of the membrane (~30 μm thick).

The simplest electrode configuration (single area) is a disk of SWCNTs concentrically applied to both sides of the membrane in equal areas, so as to produce a single uniform deformation region. More complex electrode configurations were made by patterning multiple electrodes on the same membrane, for instance, by putting multiple addressable patches one side of the membrane and a common ground electrode on the other side. We fabricated two kinds of electrode patterns: (1) a single area and (2) five-segment area. The five segment device allows for the creation and control of different strain fields, including radial expansion and contraction strains in the center, uniaxial strain in both the x and y directions, as well as rigid body displacements in both directions.

A mask for producing the electrode pattern was created by cutting (using a computer numerically controlled cutter) out a stencil of the desired electrode configuration in a PET (polyethylene terephthalate) film coated with nonstick silicone. This mask was first applied over the elastomer membrane. A thin, uniform layer of SWCNTs was prepared by vacuum filtration-transfer method, in which a water dispersion of SWCNTs was passed through a Teflon filter, depositing a uniform mat of SWCNTs[26]. The SWCNT mat was then transferred over the mask and membrane by pressing firmly and peeling off the filter and electrode mask, leaving behind the SWCNTs adhered to the membrane.

In order to transfer the metasurface to the DEA, the metasurface and a scaffold membrane (also VHB) were first bonded together, after which the scaffold membrane was bonded to the DEA. We originally thought that the adhesion between the metasurface and membrane would be enhanced by cleaning the surface using an oxygen plasma treatment to expose dangling bonds in the form of reactive hydroxyl groups. However, we found that the opposite was true: the same dangling bonds also increased the hydrophilicity, which allowed the presence of water or humidity to infiltrate and undercut the bonding between metasurface and membrane, rendering weak adhesion. Instead, pure isopropyl alcohol (IPA) was used to clean the surface of the metasurface and a chemical adhesion promoter (AP115, 3M) was used to make the metasurface slightly hydrophobic through surface silanization. AP115 was sprayed onto the metasurface and was quickly wiped off using a lint free cloth to reduce the contact time of the sacrificial layer with the small amount of water present in AP115. Although we do not know the exact chemical contents of AP115, the active ingredient is probably 3-glycidoxypropyltrimethoxysilane. Next, the membrane was pressed onto the metasurface, using a smooth, spherical press made of soft silicone. A spherical press was important to achieve good contact throughout the entire metasurface area, by overcoming the nanoimprint proximity effect[30], by which there is less adhesion in the center than the periphery when using a flat press. The entire sample was then heated in an oven at 50 °C for 3 hours to improve adhesion by allowing the membrane polymer to flow and then cooled to room temperature.

For release, the entire sample was immersed in water. Dissolution began around the edges of the devices and moved towards the center as water slowly percolated between the nano-posts. The release time of the entire device depended on the device area and varied between minutes to hours. The scaffold membrane with the supported metasurface was then attached to the DEA by gently pressing. We refer to the scaffold membrane after combination as the intermediate elastomer layer (IEL). Two types of layered devices were made, which we refer to as single layer (SL) and double layer (DL) devices (Fig. 3a): the SL device contains an IEL (not pre-stretched) which was cut to be as small as the metasurface, while the DL device contains an intermediate elastomer layer (pre-stretched 4x, linear) of area equal to the entire DEA, effectively increasing the stiffness of the combination.

In the final step, electrical contacts were made to the CNT electrodes. The SWCNT electrode patches were connected by conductive silver grease to leads of conductive carbon tape, which were then connected to wires.

**Measurement methods.** A tunable laser (HP 8168F) operating between the wavelengths of 1440-1590 nm was used as the light source. This optical output of this laser was connected to an optical fiber collimator (Thorlabs F810APC-1550), which produced a collimated beam, 7 mm in diameter. The collimated beam was used to illuminate the device. The polarization of the beam was allowed to wander because the device was designed to be polarization independent, and no polarization dependence was observed. A high voltage source (Trek 610E) was used to tune the device. After passing through the device, the light was measured by a conventional horizontal microscope setup: a microscope objective (10x Mitutoyo M Plan Apo NIR infinity corrected objective) and a tube lens (Plano-Convex Lens, f = 200 mm) was used to magnify the beam in order to fully visualize the focal spot on the camera (digital InGaAs, Raptor OWL640). The entire horizontal microscope setup was mounted on a linear motor (NPM Acculine SLP35), which allowed horizontal scanning of the light field with a positional accuracy of 1 μm. This setup (see Supplementary Fig. S4a and S4b) was used to characterize the focal length and focal spot size. To measure the efficiency, we replaced the microscope objective, tube lens, and camera with an optical power meter (Thorlabs PM100D) (see Supplementary Fig. S4c).

**Data availability.** The datasets generated during and/or analysed during the current study are available from the corresponding authors on reasonable request.

**Figure 1: Design of tunable metasurface lens.**

An illustration of a metasurface (left column) is constructed by digitizing an analog optical phase profile on a flat surface into discrete cells, each of which contains a metasurface element that locally imparts the required phase shift to the incident light in order to reconstruct the desired wavefront (middle column, dotted line is optic axis). The wavefront generated by the metasurface determines the subsequent beam shaping (right column). A periodic, radial gradient structure can be seen in the metasurface illustration, corresponding to 0 to 2π phase modulation, similar to the rings of a Fresnel lens. Rows: **(a)** A metasurface with a hyperboloidal phase profile following Eq. 1 will focus normal incidence light to a point at distance corresponding to the focal length. Light rays traveling along the x, z and y, z planes (red and blue lines, respectively) will both be focused to the same point. **(b)** When the metasurface is stretched uniformly and isotropically, the phase profile widens and shallows, and this is seen in the outgoing wavefront (i.e., isophase surface), resulting in an extended focal length. Conversely, if the metasurface is compressed, the focal length would reduce. **(c)** When the metasurface is asymmetrically stretched (here stretched and compressed in the x and y directions, respectively), astigmatism is produced, such that light rays traveling along the x, z and y, z planes experience different focal lengths. **(d)** When the metasurface is displaced laterally in the x, y plane, the entire phase profile and hence focus position is correspondingly laterally shifted.

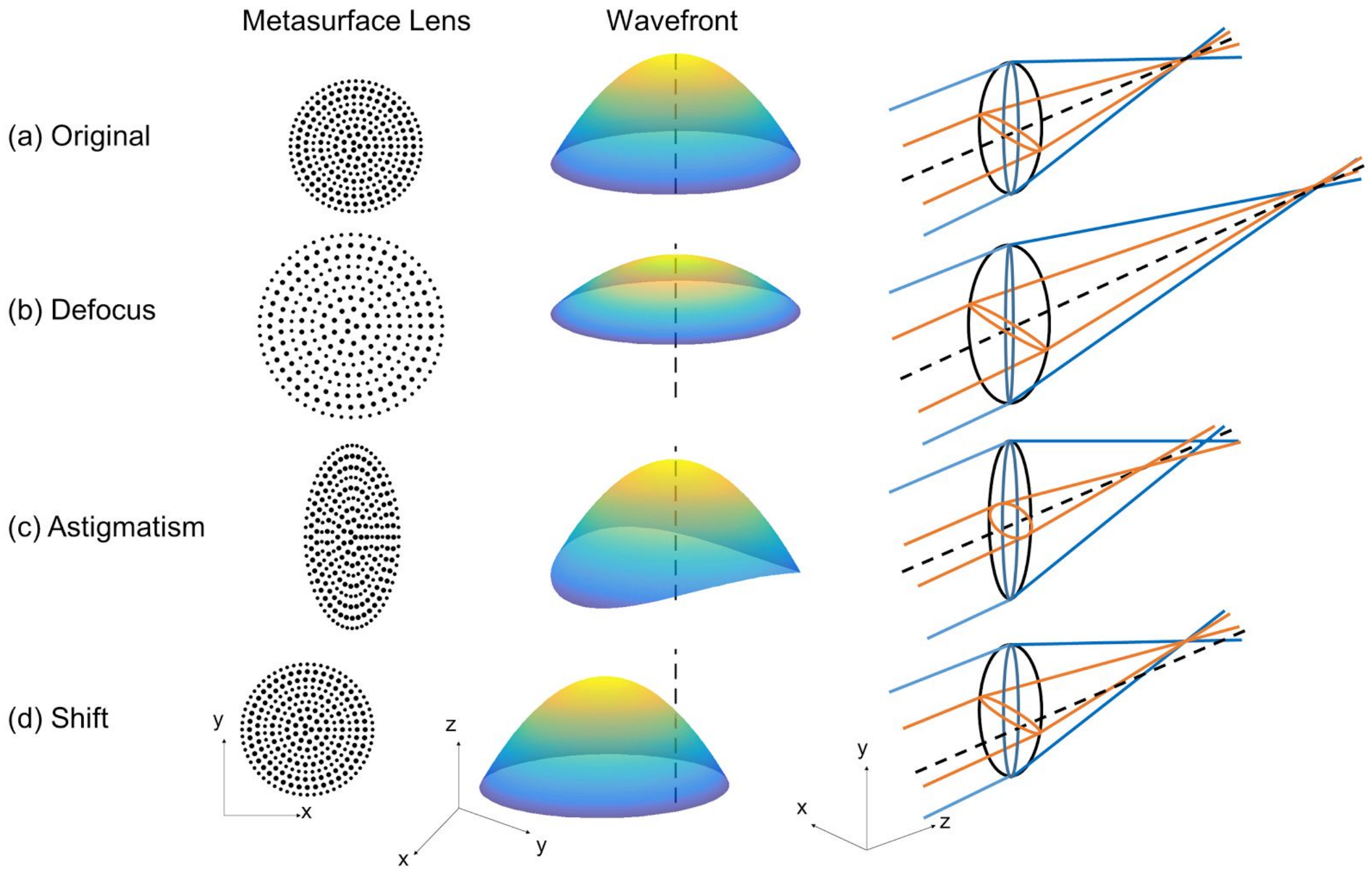

**Figure 2: Dielectric elastomer actuator platform.**

**(a)** Fabrication steps: (i) Schematic of the thin-film stack prepared for photolithography. (ii) After patterning and etching, the stack is attached to an elastomer layer and the whole stack is immersed in a water bath. (iii) Schematic of the release process showing the dissolution of the sacrificial layer ($GeO_2$) from the outer edge of the device towards the center, leaving the metasurface structure attached to, and supported by, the elastomer layer. (iv) A schematic of metasurface and DEA combination in which an applied voltage supplies the electrode layers (SWCNTs) with electrical charges. Their electrostatic attraction compresses the elastomer in the thickness direction, and causes expansion in the lateral direction. **(b)** False-colored scanning electron microscope image of the center of the lens shows a-Si posts on the $GeO_2$ sacrificial layer before attaching to the elastomer layer.

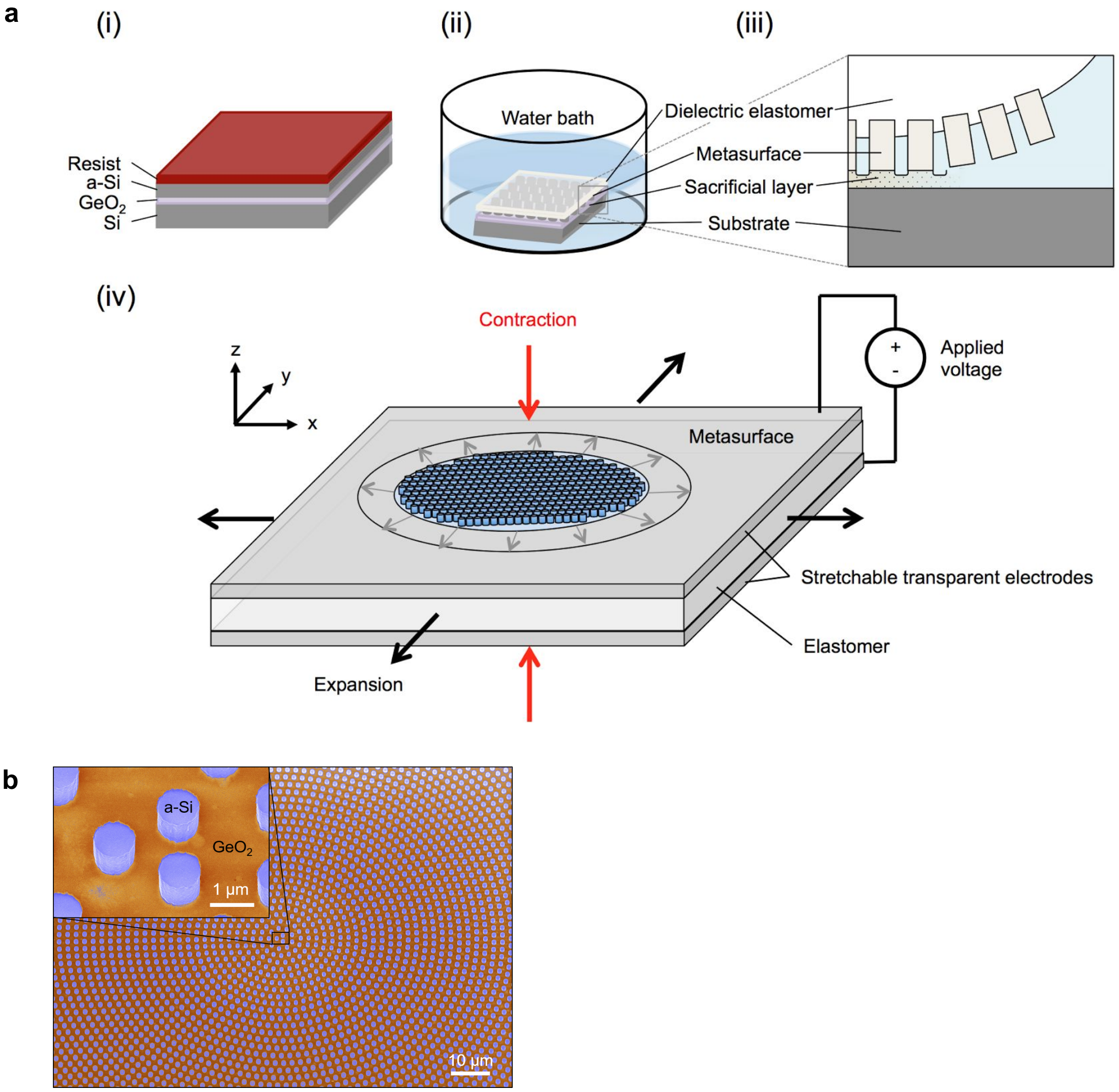

**Figure 3: Device design.**

**(a)** Top left: The device with the metasurface at the center of a pre-stretched DEA sheet, suspended by a plastic frame. The electrical contacts are made to the SWCNT electrodes using silver grease and carbon tape (black stripes). There are eight connections using SWCNT, labeled V1-5 and GND (ground). Top right: Top view of the five electrode configuration. Bottom: Schematics (cross section) of single layer (SL) and double layer (DL) devices. Both devices have the same basic structure in which the metasurface is attached to a DEA via an intermediate elastomer layer (IEL). In the SL, the IEL is only as large as the metasurface, while in the DL device, it covers the entire DEA. **(b)** Different electrode activation configurations are shown. In general, applying a voltage to an electrode will cause the area under that electrode to expand. The defocus (+) tuning is controlled by the center electrode. The defocus (-) tuning is controlled by applying voltages of equal magnitude to all four peripheral electrodes. Vertical astigmatism (X) is controlled by the left and right pair of peripheral electrodes and horizontal astigmatism (Y) is controlled by the top and bottom pair of peripheral electrodes. The X shift (+) and (-) is controlled by the left and right peripheral electrodes, respectively. The Y shift (+) and (-) is controlled by the bottom and top peripheral electrodes, respectively.

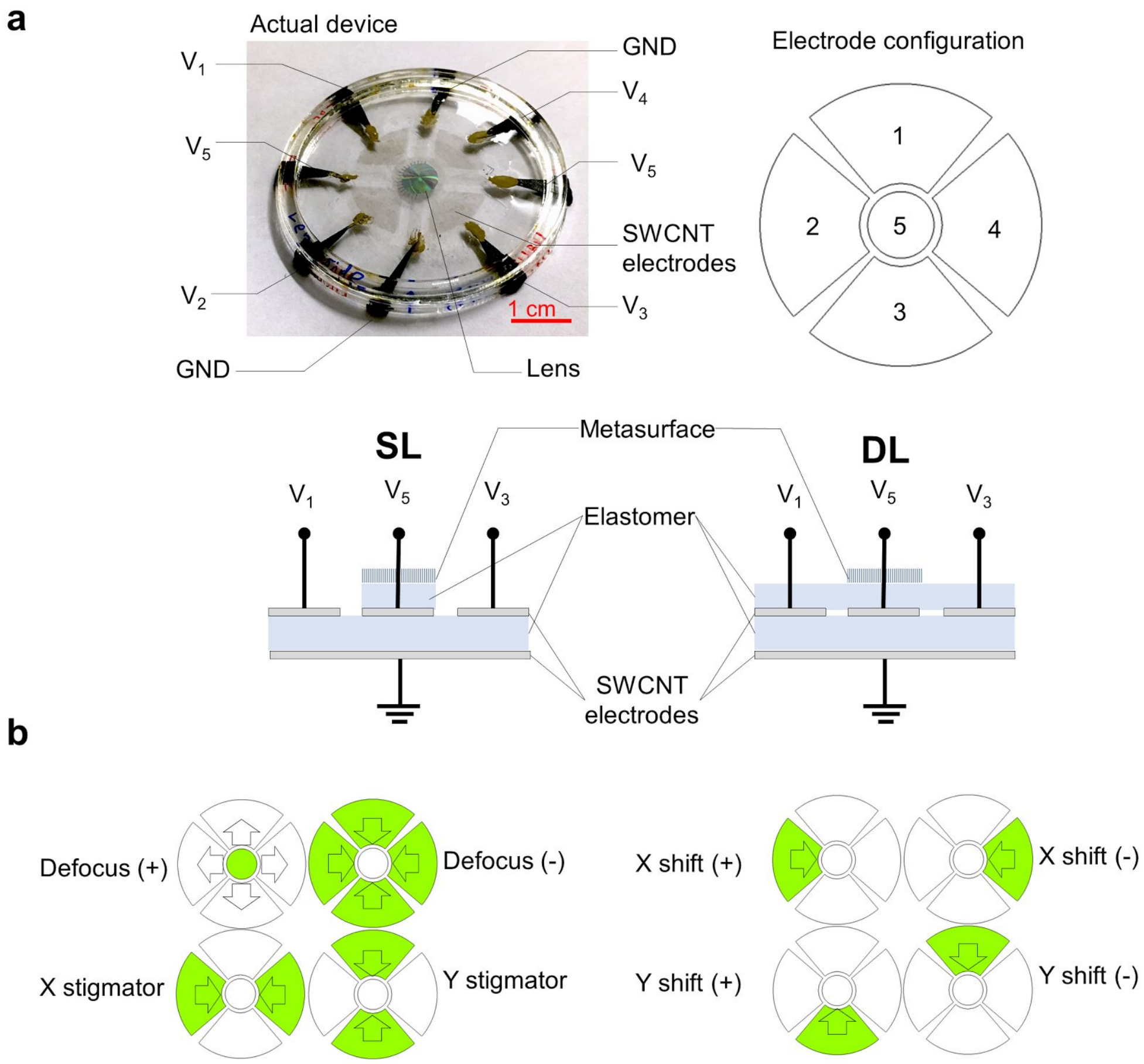

**Figure 4: Measurement results.**

**(a)** Optical microscope images (top row, scale bar: 20 μm) and their corresponding Fourier transforms (bottom row) at the center of the metasurface lens on the DE platform at (i) no voltage, (ii) 2.5 kV applied to the center electrode, and (iii) 2.75 kV applied to tune X-astigmatism. The dark spots are defects introduced during the transfer process. They are either missing or tilted posts. **(b)** Measurement of focal length tuning using center electrode $V_5$. Blue dots with error bars (standard deviation): Optical measurement of device focal length as a function of applied voltage. Solid blue line: fit focal length data to equation 3, showing good agreement ($R^2$ = 0.9915). Red triangles: Measurement of stretch as a function of the applied voltage. **(c)** Measured focusing efficiency with error bars (standard deviation) as the voltage is varied (measurement description, see Supplementary Fig. S4c). **(d)** The focus profile was scanned using a microscope objective (10x magnification) at two different voltages: 0 kV (top) and 1 kV (bottom). (Left) Z-scan of the intensity profile showing two distinct focal lengths. (Center) Image of focal spot intensity profile, i.e. x-y cross sections at the position of maximum intensity. (Right) Line scans of focal spot intensity image in the x (blue) and y (red) directions in comparison to the theoretical diffraction-limited spot size (black). Bottom: Also shown are the images of the focal spots at 0 through 2.5 kV captured directly by the camera without the microscope objective (scale bar: 500 μm). **(e)** As the applied voltage is tuned using the x-stigmator electrode configuration, the measured Zernike coefficients of the phase profile are plotted, showing targeted tuning of vertical astigmatism, while other Zernike coefficients exhibit little change throughout the tuning range. The defocus value accounts for the original focusing power of the lens. **(f)** Measurement of X,Y-shift control as 1.9 kV applied, showing two-axis control over displacement from the focus position (center red circle) at 0 kV.

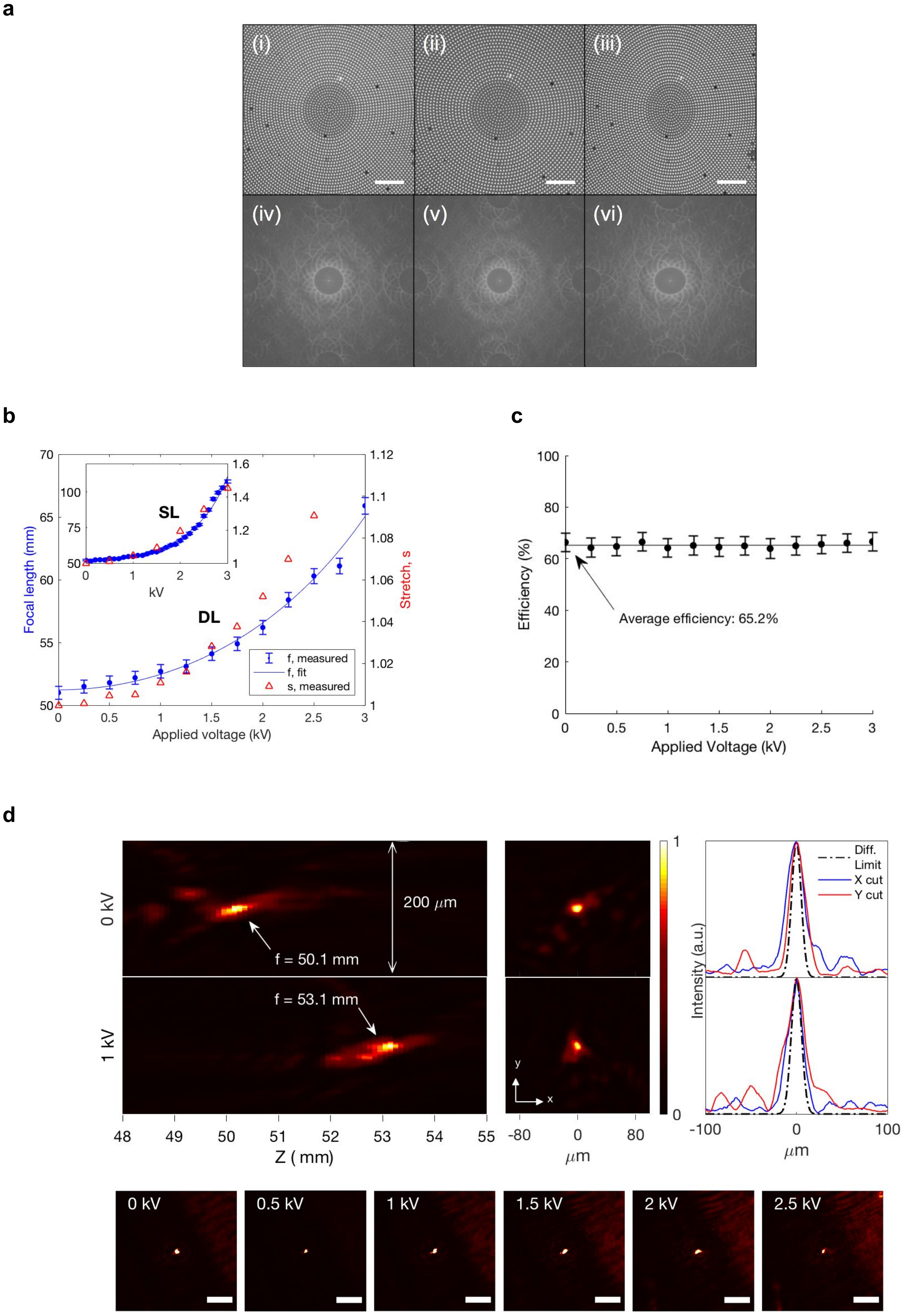

**e**

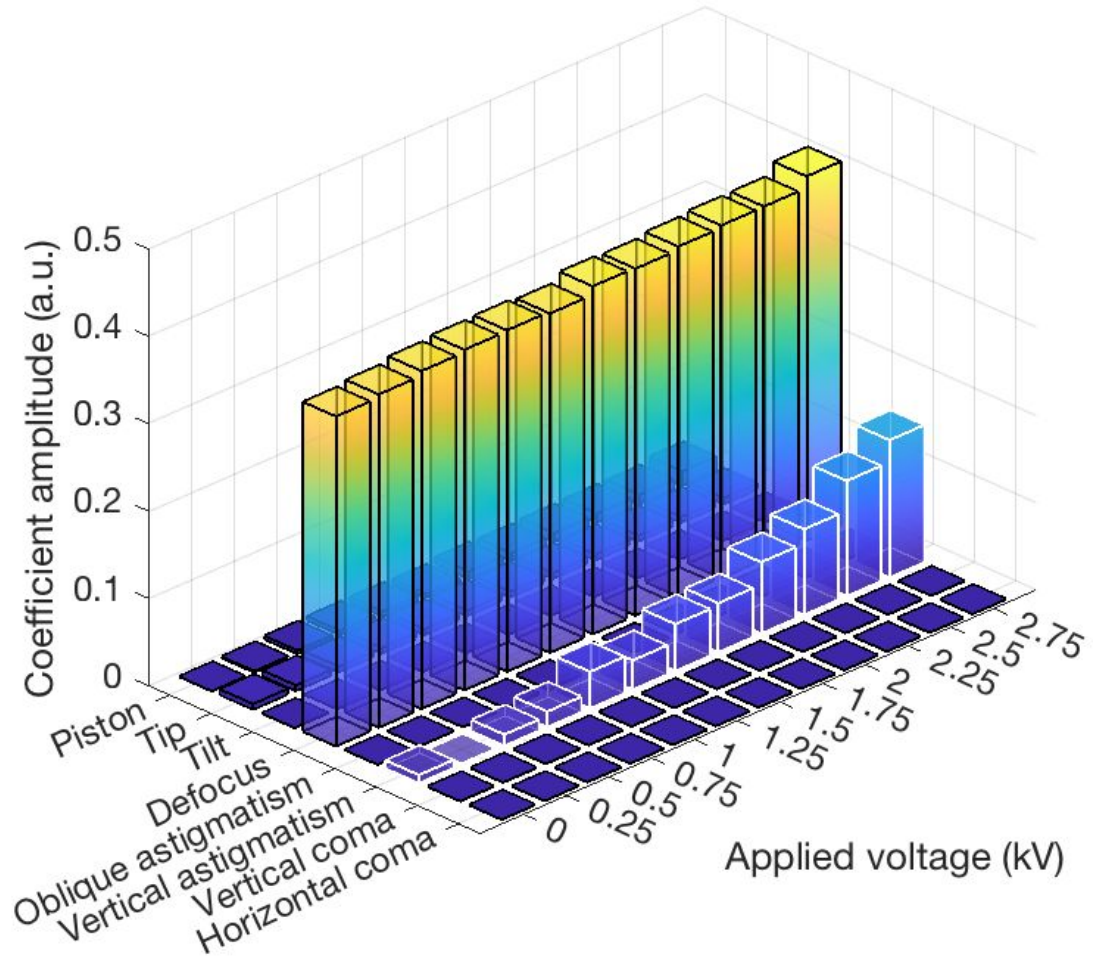

**f**

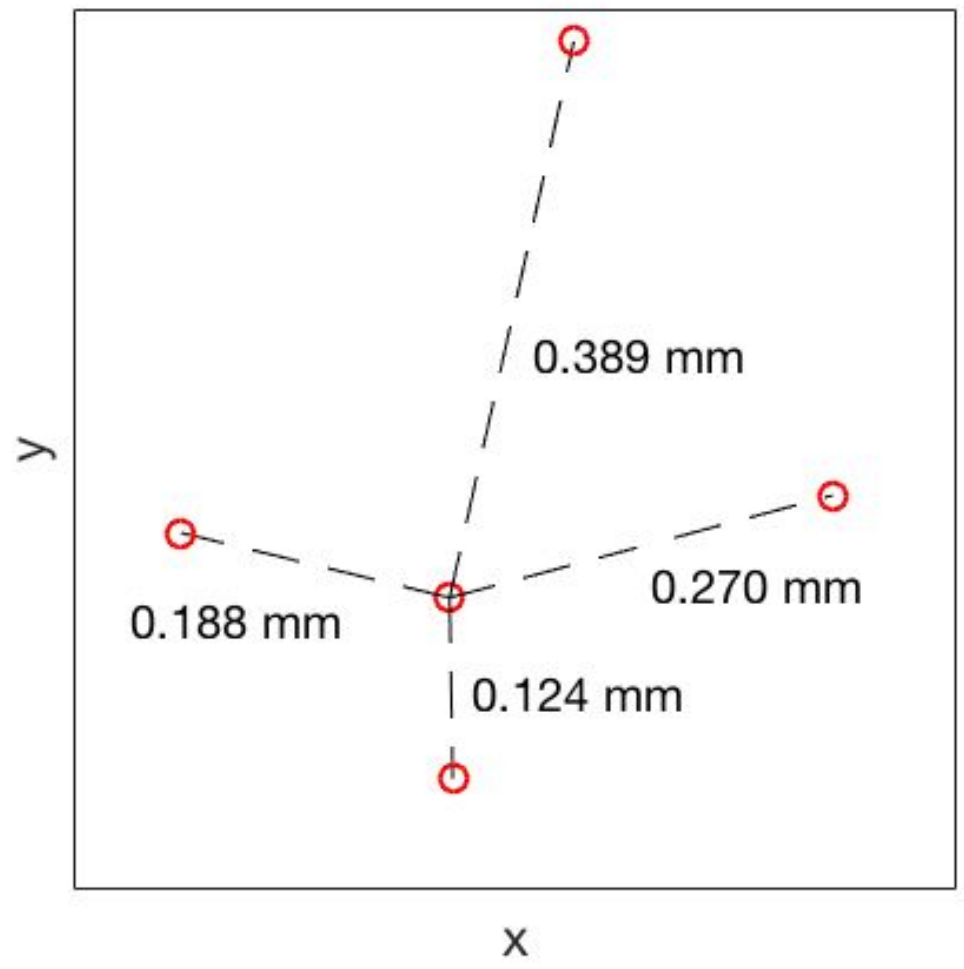

## Supplementary Information

### Uniform stretching approximation

A lens that focuses normal incidence light has the following hyperboloidal phase profile (Eq. 3):

$$\phi_{ideal} = \pm k\left(\sqrt{r^2 + (s^2 f_0)} - s^2 f_0\right),$$

where we desire a tunable focal length relationship of the form $f = s^2 f_0$.

In order to achieve $\phi_{ideal}$, a stretch profile that depends on the radial position is required: $s_{radial}(r)$. Then the phase profile of a lens with focal length $f_0$ that is stretched by $s_{radial}$ undergoes the coordinate transformation: $r \rightarrow r/s_{radial}(r)$, such that the resulting phase profile is the following:

$$\phi_{radial} = \pm k\left(\sqrt{(r/s_{radial}(r))^2 + {f_0}^2} - f_0\right).$$

In order to find the stretch profile that produces the target phase profile $\phi_{ideal}$, which has a nominal, uniform stretch value $s$, we can equate $\phi_{radial} = \phi_{ideal}$ and write $s_{ideal}$ in place of $s_{radial}$ (where $s_{ideal}$ is a specific case of $s_{radial}$ that produces $\phi_{ideal}$):

$$\sqrt{(r/s_{ideal}(r))^2 + {f_0}^2} - f_0 = \sqrt{r^2 + (s^2 f_0)^2} - s^2 f_0.$$

Solving for $s_{ideal}$, we find the following:

$$s_{ideal}(r) = r\left(r^2 + 2(s^4 - s^2){f_0}^2 + 2(1 - s^2) f_0 \sqrt{r^2 + (s^2 f_0)^2}\right)^{-1/2}.$$

To lowest order, we see that

$$s_{ideal} \approx s.$$

Although $s_{ideal}$ is not exactly equal to $s$, their difference is extremely small. Fig. S1a and S1b show that there is a slight decrease in local stretch as the radial position is moved outwards from the center of the lens, using the lens parameters of our experiment ($f_0 = 50\ mm$). A nominal stretch of $s = 1.2$ yields a difference of only $\Delta s = 1 \times 10^{-4}$ at a radial position of 3 mm, corresponding to the size of the lens in our device.

### Focal length vs voltage relation

Volume conservation of the elastomer requires that: $s_x s_y s_z = 1$, where s is the stretch. For isotropic materials under uniaxial compression: $s_x = s_y = s$, where $s = 1 + \varepsilon_L$ ($\varepsilon_L$ is the lateral strain) and $s_z = 1 + \varepsilon_z$ ($\varepsilon_z$ is the longitudinal strain, i.e. in the thickness direction), giving

$$s^2 = 1/s_z.$$

For small strain ($\varepsilon_z \lesssim 0.2$), the strain response $\varepsilon_z$ of materials under Maxwell's compression stress, $\sigma_M = -\epsilon E^2$, can be approximated as that of a linear elastic material with Young's Modulus of Y:

$$\varepsilon_z = -\epsilon E^2 / Y \ ,$$

where E is the electric field and $\varepsilon$ is permittivity. As $s_z = 1 + \varepsilon_z$ and $f/f_o = s^2$, the relation of focal length with voltage, V, is the following (Eq. 4):

$$\frac{f}{f_0} = \frac{1}{1-(\varepsilon/Y)(V/t)^2} \ ,$$

where *t* is the instantaneous thickness of the elastomer, which is a function of the applied voltage, but can be approximated as constant since it changes very little[1]. By writing $b = \varepsilon/(Y t^2)$, the series expansion for small voltages near V=0 gives

$$f/f_0 = 1/(1 - bV^2) = 1 + bV^2 + b^2 V^4 + O(V^6) \ .$$

We see that the focal length relation transitions from a quadratic relation ($V^2$) to a quartic relation ($V^4$), as the voltage is increased, $V_{transition} > b^{-1/2}$ (and to higher orders if V is increased further). For example, by taking nominal values for our dielectric layer made of VHB: $\epsilon_r$ = 6, Y = 1.8 MPa, and t = 30 μm, $V_{transition}$ is 5.52 kV.

**Figure S1: Uniform stretch approximation.**

**(a)** The ideal stretch profile ($s_{ideal}$, which is the stretch profile that produces an aberration-free hyperboloidal phase profile) and the nominal, uniform stretch value ($s$) is plotted for typical values of $s$ from 1.01 to 1.3 (they are the same for $s = 1$) as a function of radial position. **(b)** The difference $\Delta s = s_{ideal} - s$ is plotted for typical values of $s$ from 1 to 1.3 (blue to red lines, respectively), as a function of radial position for an example lens with focal length of 50 mm. **(c)** The magnitude of the wavefront aberration function (WAF) in radians (color bar), here defined as the difference between stretched ($\phi_{stretched}$) and target ($\phi_{target}$) phase profiles, is plotted as a function of radial position ($r$) and stretch ($s$) for a lens with focal length of 50 mm using light of wavelength 1550 nm. The white, dotted line shows the maximum aberration, which occurs at a stretch of $s \approx 1.22$.

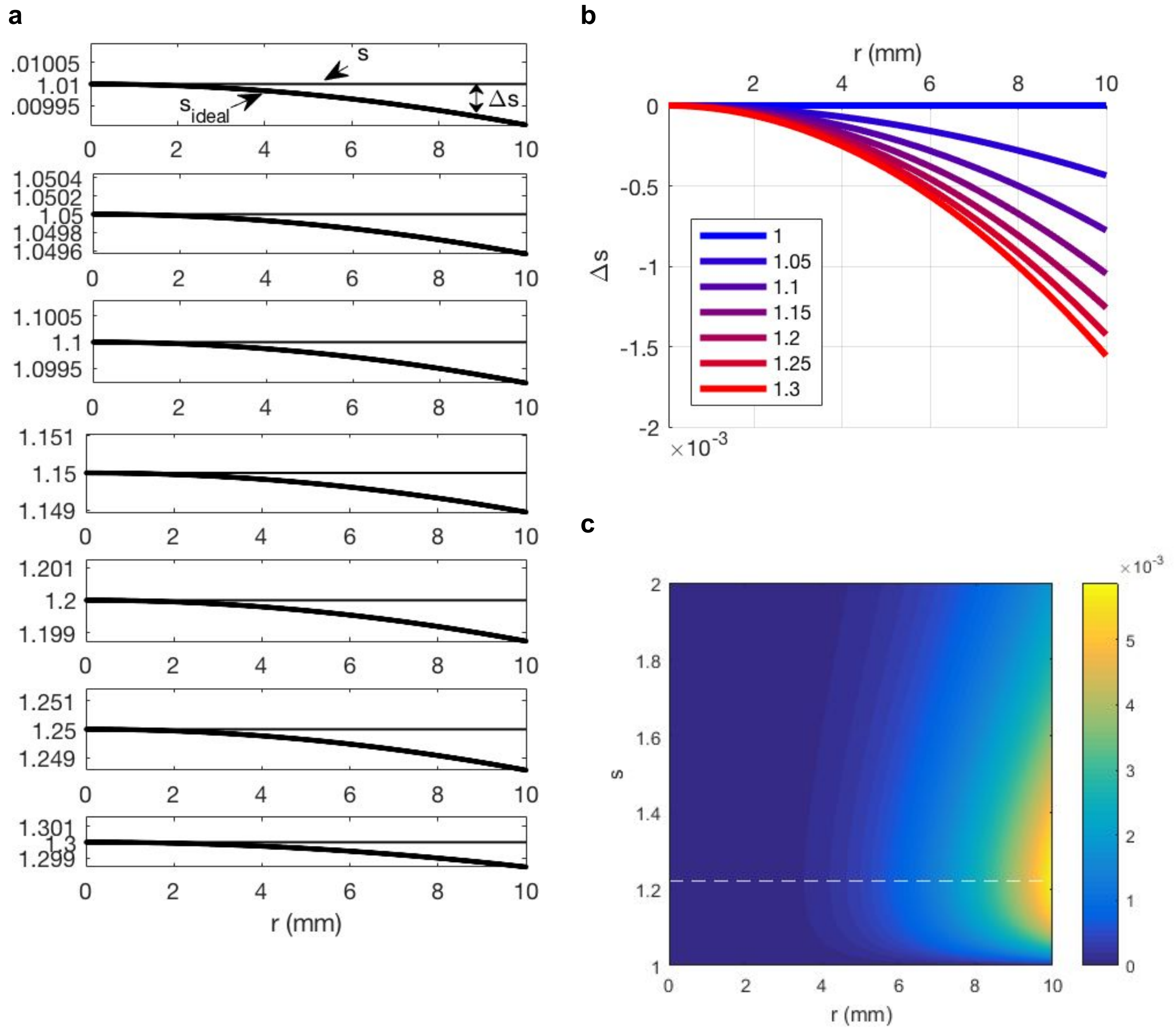

**Figure S2: Metasurface lens design elements.**

**(a)** The unit cell of the lens consisting of the $SiO_2$ substrate and an amorphous silicon (a-Si) post. The dimension are labelled as follows: h is the height of the post, d is the diameter of the post, u is the size of the unit cell and e is the constant edge-to-edge spacing between neighboring posts. **(b)** Phase and amplitude response of posts are plotted as a function post diameter. **(c)** The metasurface design of our device is shown. Left: The whole lens is 2 cm in diameter and appears as a solid black circle in this rendering due to the sheer density of structures. Top right: a close up view showing a 200 μm wide window of the center of the lens. Bottom right: a close up view showing a 50 μm wide window of the center of the lens.

**a**

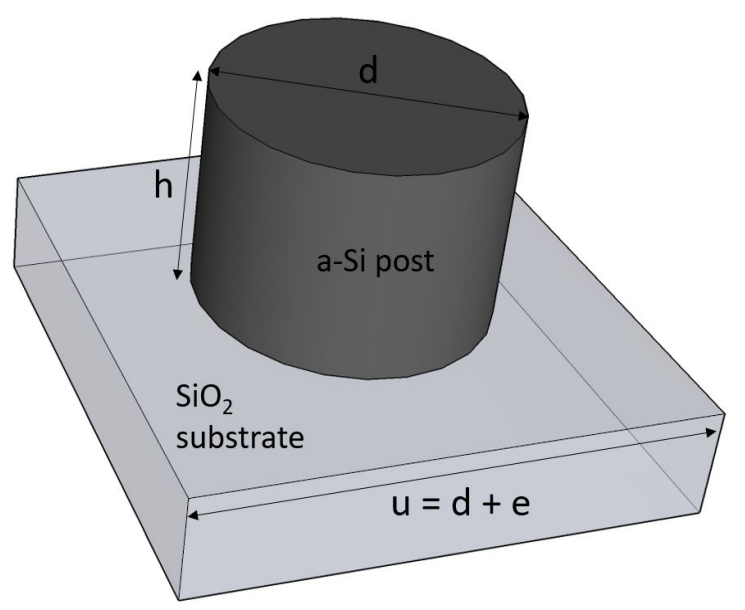

**b**

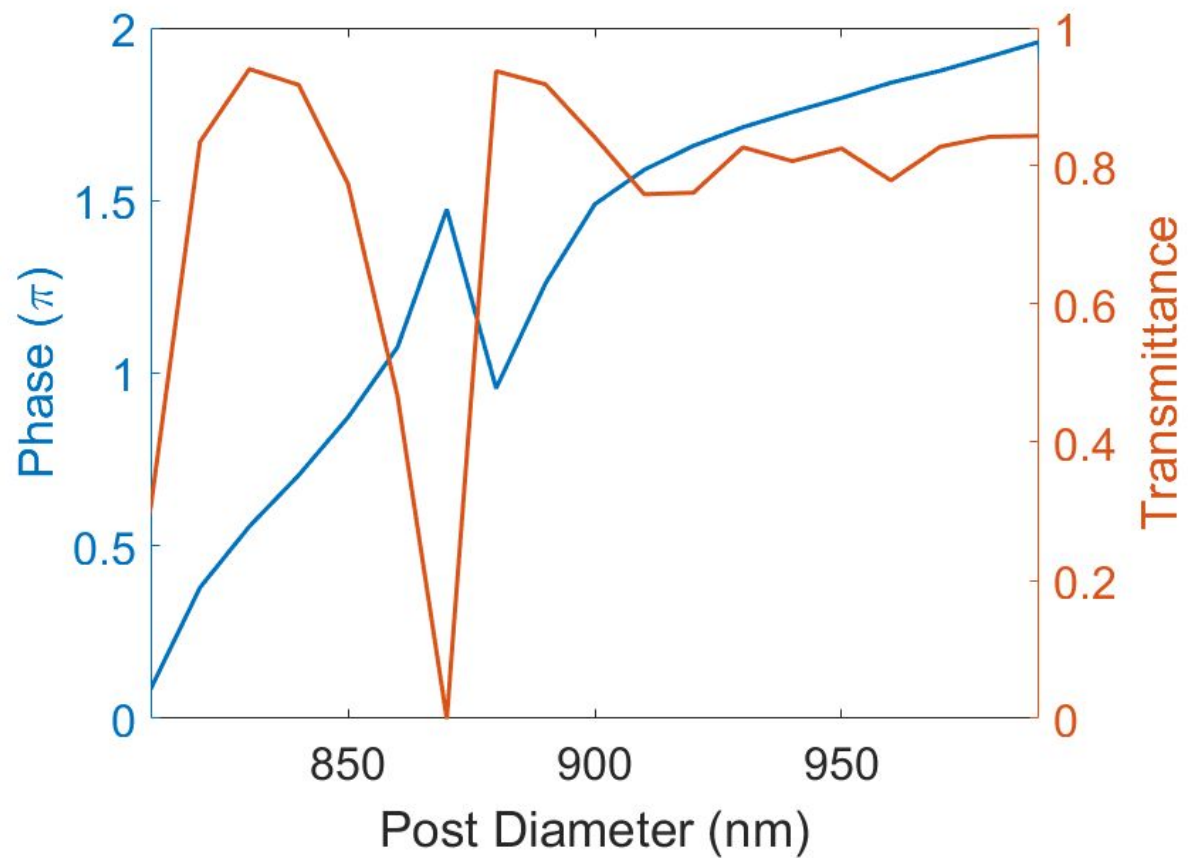

**c**

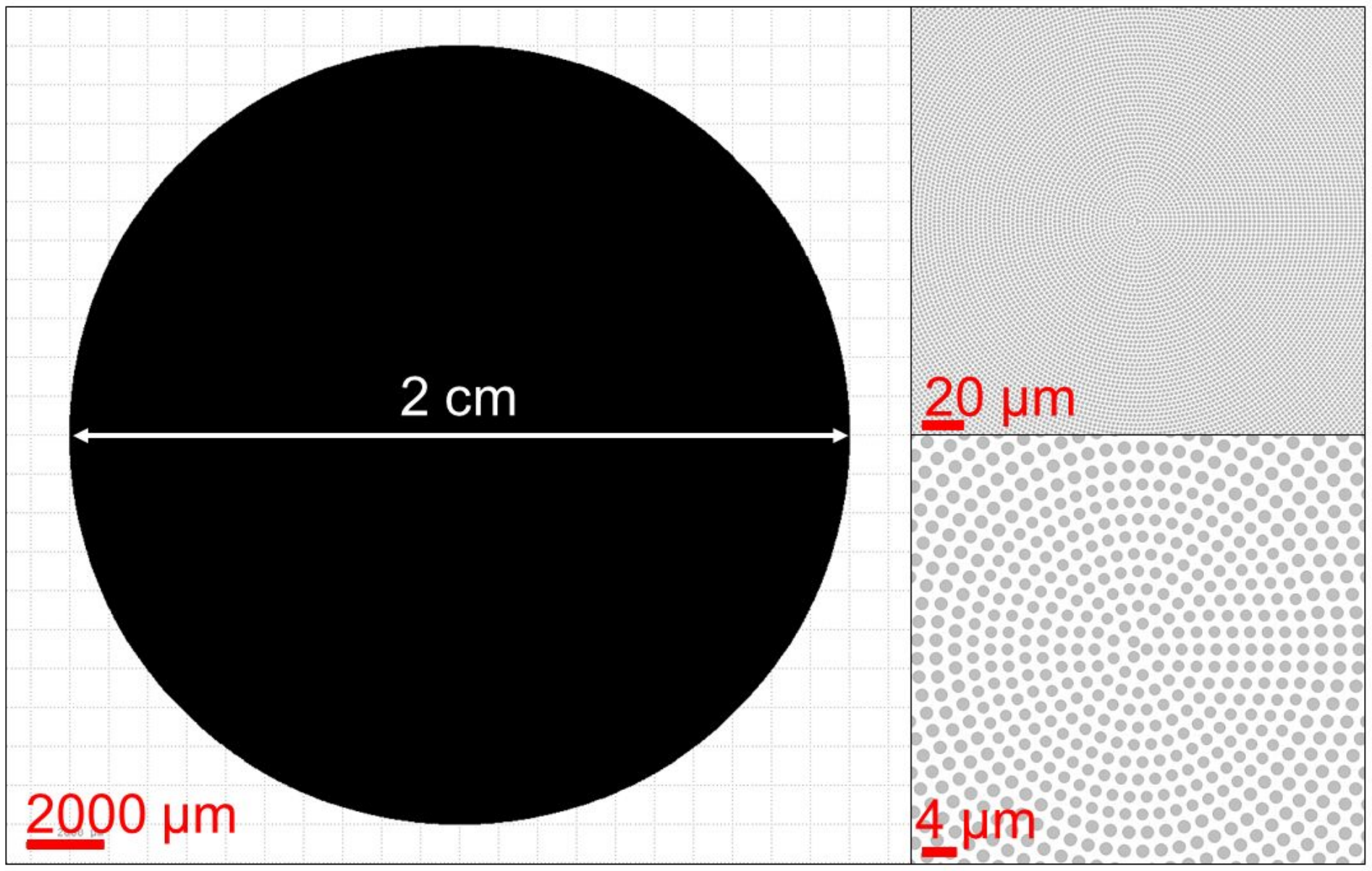

**Figure S3: Optical measurements of DEA, consisting of pre-stretched (4x) acrylic elastomer (VHB 4905, 3M) and single-walled carbon nanotube (SWCNT) electrodes.**

The measurements were performed using the Agilent Cary 7000 Universal Measurement Spectrometer and the Bruker Lumos FTIR microscope. The spectrum range is from 200 nm to 16 µm. The x-axis of the plots is in log scale. Our device operates at 1550 nm. Top: Transmission (blue), reflection (green), and absorption (yellow) measurements for a single VHB membrane layer. Middle: Transmission, reflection and absorption measurements for a device with the VHB layer and single-walled carbon nanotube (SWCNT) electrodes applied to both sides. Bottom: Derived absorption spectrum of SWCNT layers only. At 1550 nm, the absorption is close to zero.

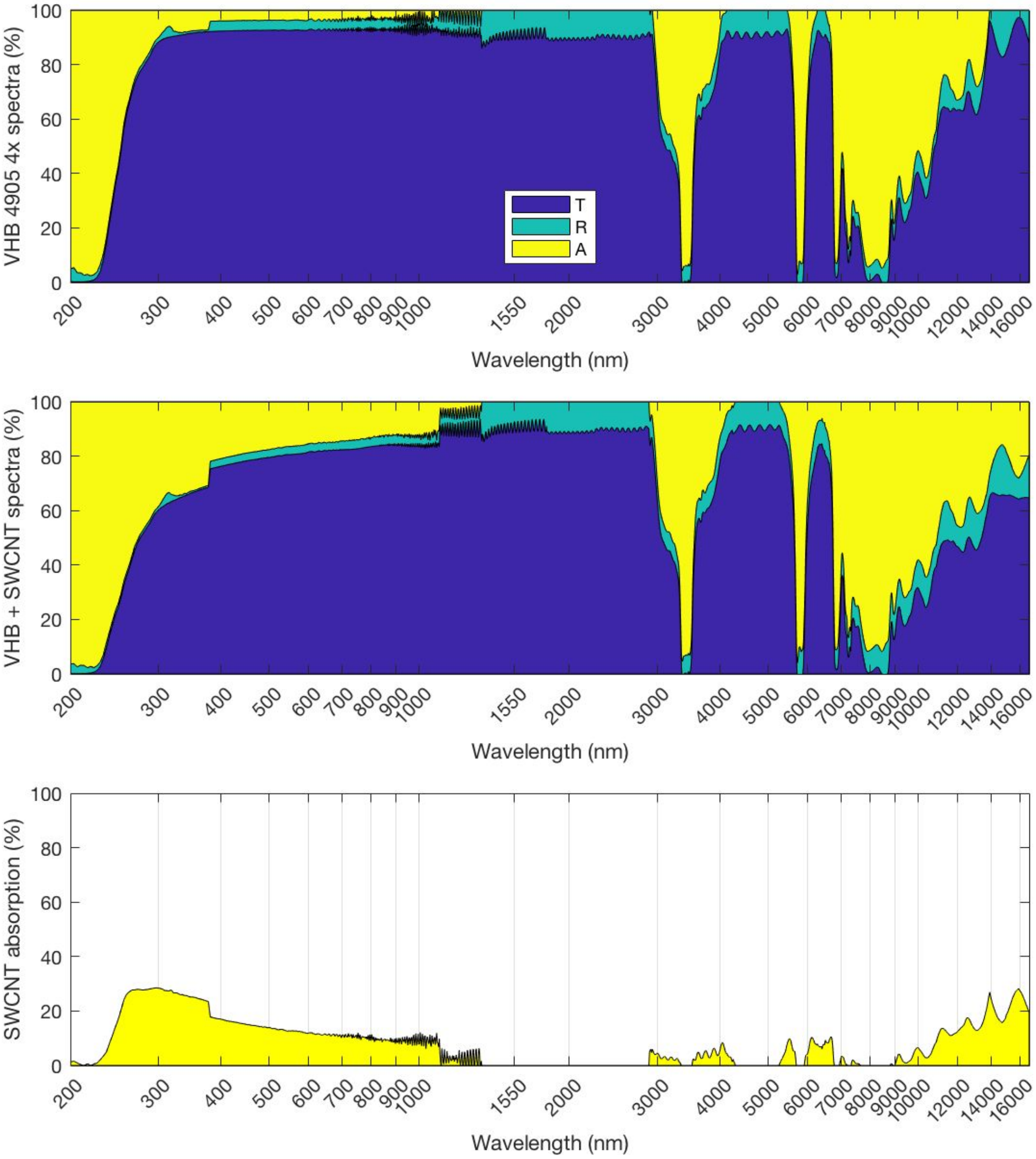

**Figure S4: Measurement setup.**

Schematics (top view) of the setups for measuring the focused spot size of the device **(a)** with **(b)** without the horizontal microscope to magnify the image. The setup without the horizontal microscope was used in at higher applied voltages, in which there was not enough space on the motorized stage to accommodate the horizontal microscope given the extended focal length of the device. **(c)** Setup for measuring the focusing efficiency of the device. The first aperture was used to reduce the incident beam size to 6 mm to match with the size of the device. The second aperture was used to exclude the stray light. The focusing efficiency was calculated using the ratio of the optical power at the focus (measured with both apertures and the device in place) to the total unobstructed optical power (measured with both the device and second aperture removed, but with the first aperture still in place).

**a**

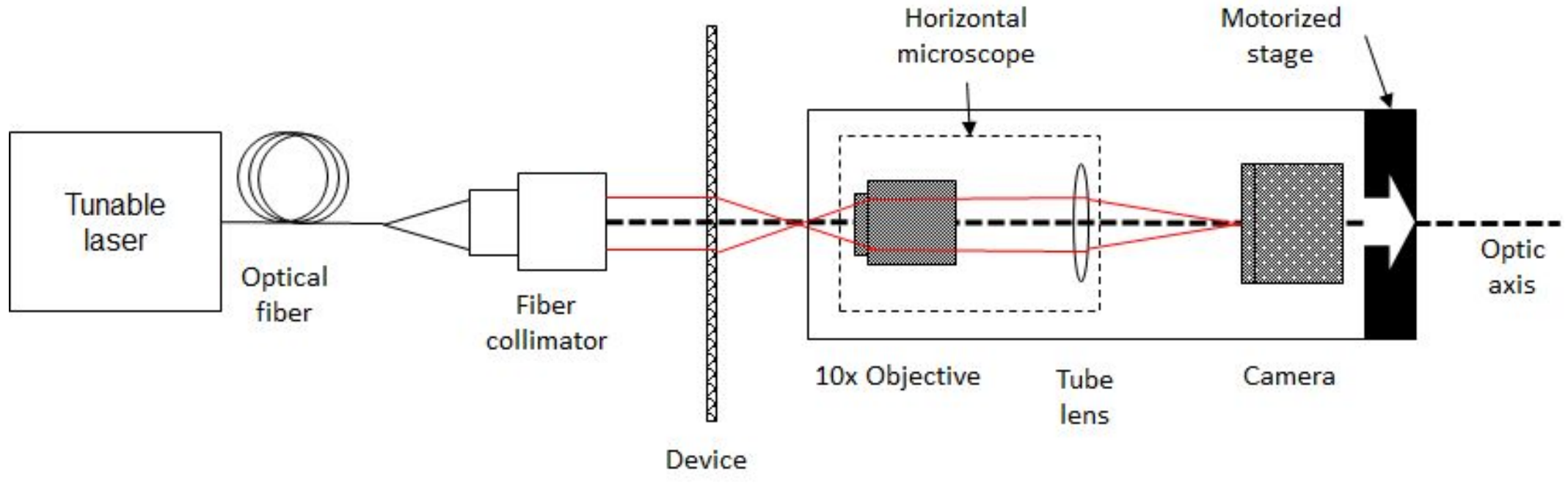

**b**

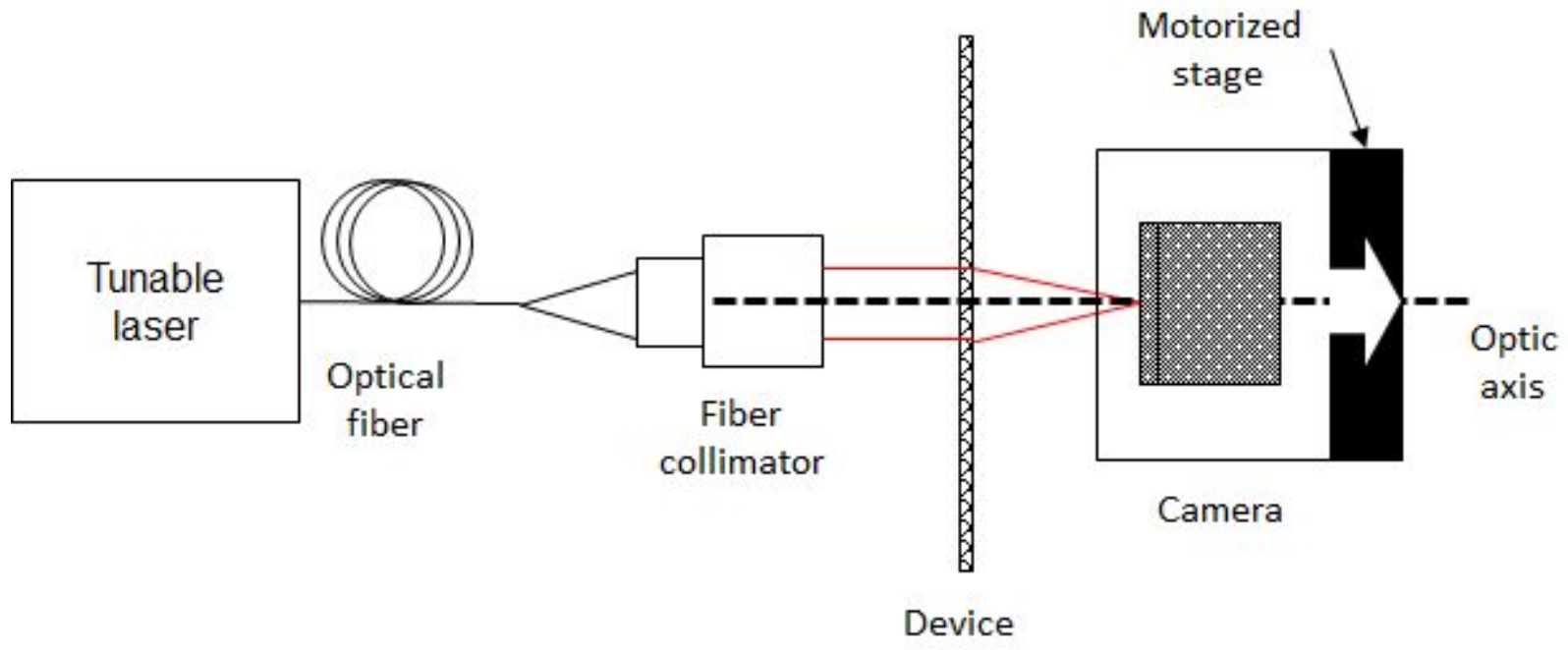

**c**

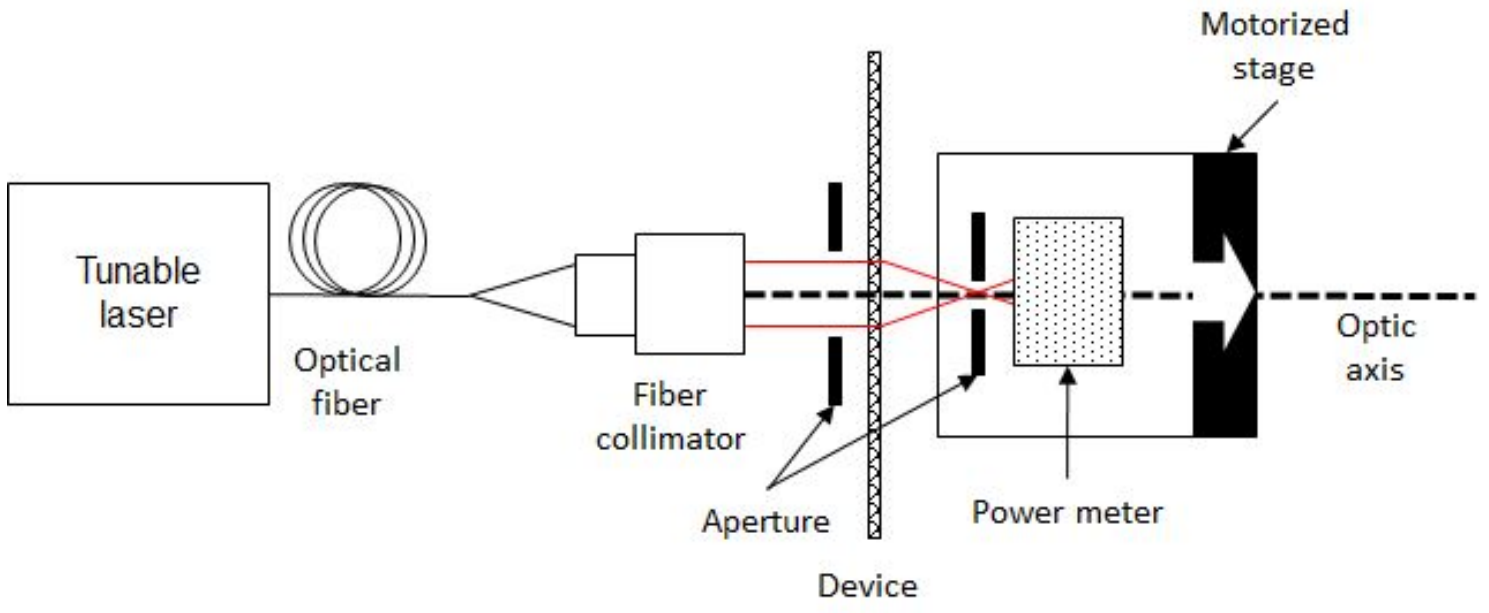

**Figure S5: XY-shift distortion measurement.**

XY-shift distortion was measured by using microscope images of metasurfaces, for which displacements were produced in the up, down, left, right ((i)-(iv), respectively, with scale bar: 20 μm) directions by applying 1.9 kV to the four peripheral electrode areas. Shown below ((v)-(viii)) are the corresponding Fourier transforms of (i)-(iv), respectively. During imaging, the microscope field of view was re-adjusted with respect to the center of the metasurface after each voltage application.

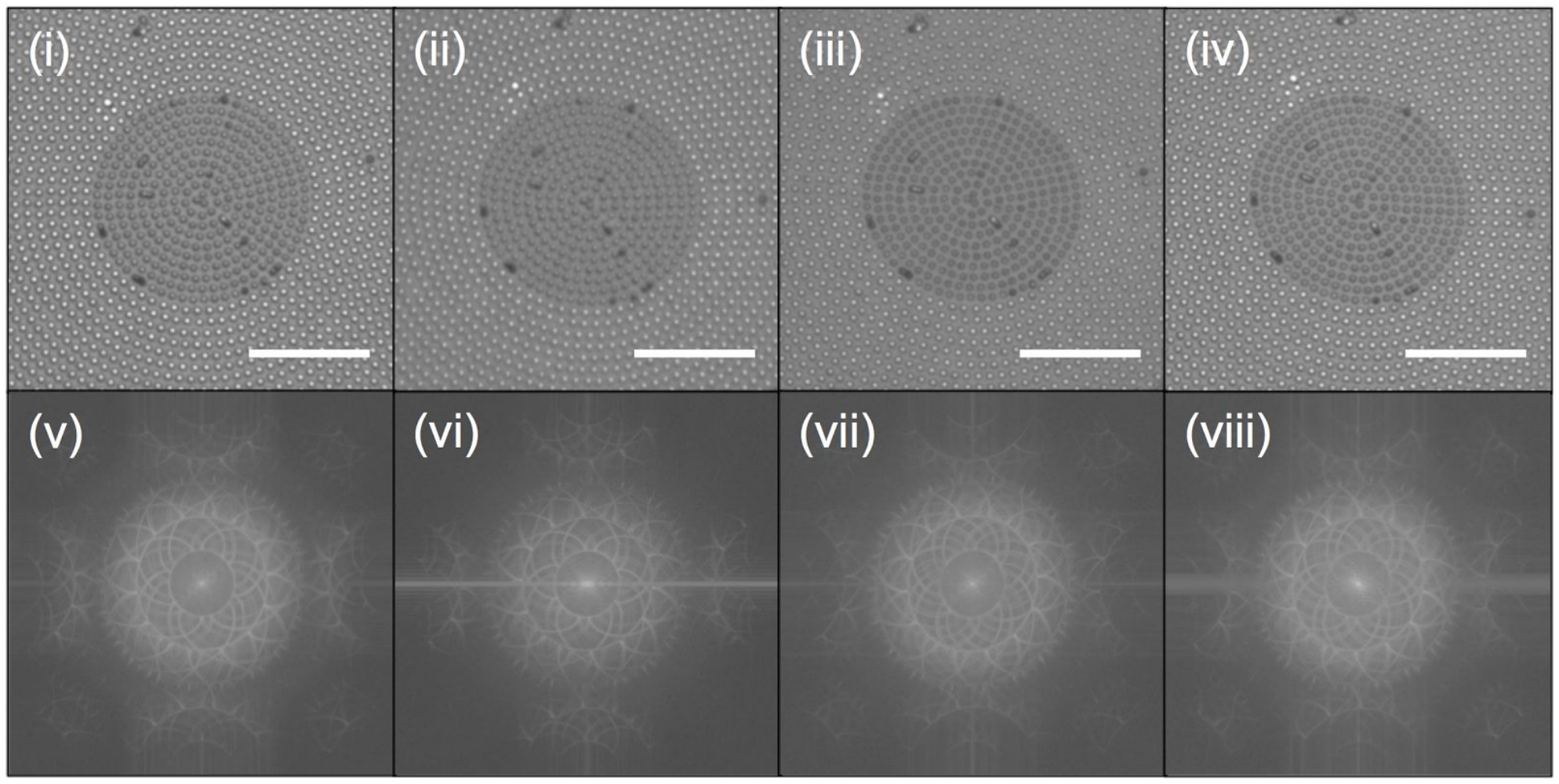

**Figure S6. Fitting measurement of focal length tuning using electrode $V_5$.**

Fitting measured focal length with applied voltage for **(a)** SL device and **(b)** DL device. Black dots: Measurement of device focal length as a function of applied voltage. Blue solid lines: Fit of the relationship between the focal length and the applied voltage. Blue dashed lines: Fits with 95% confidence bounds.

**a**

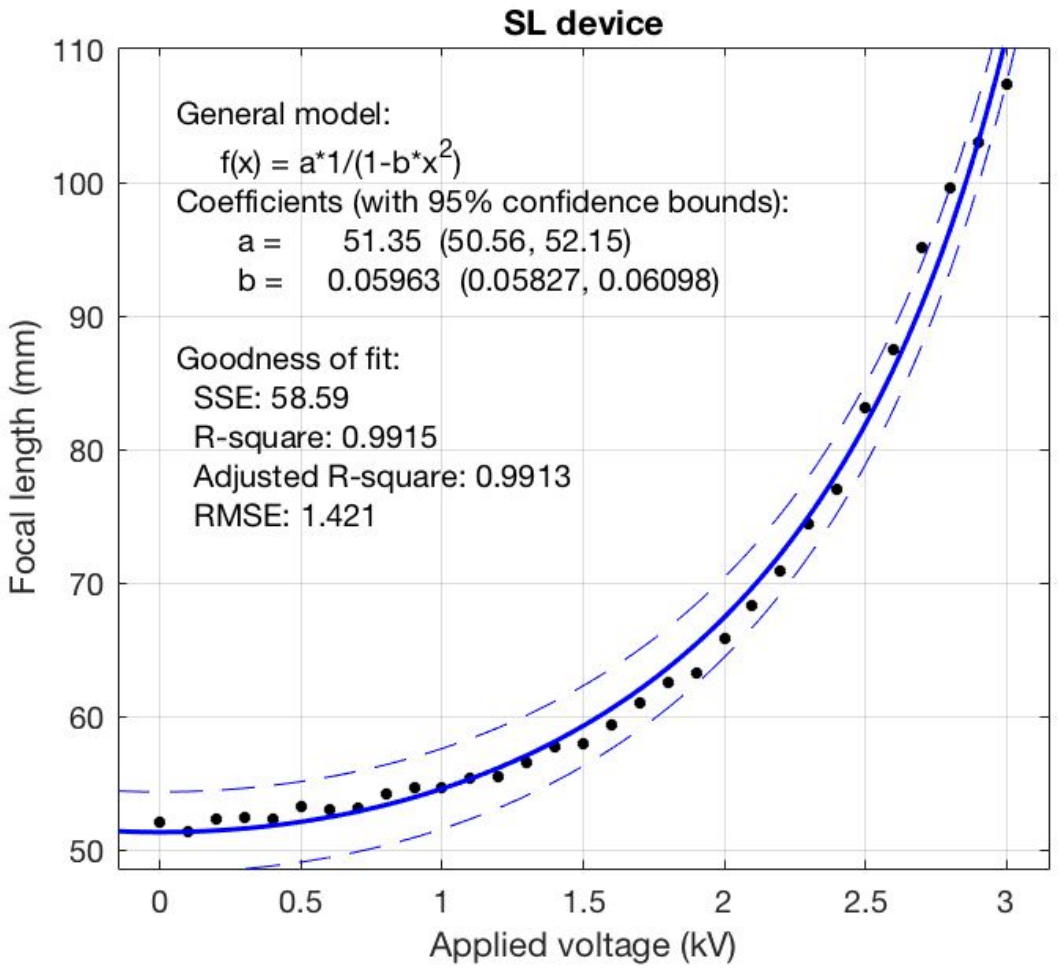

**b**

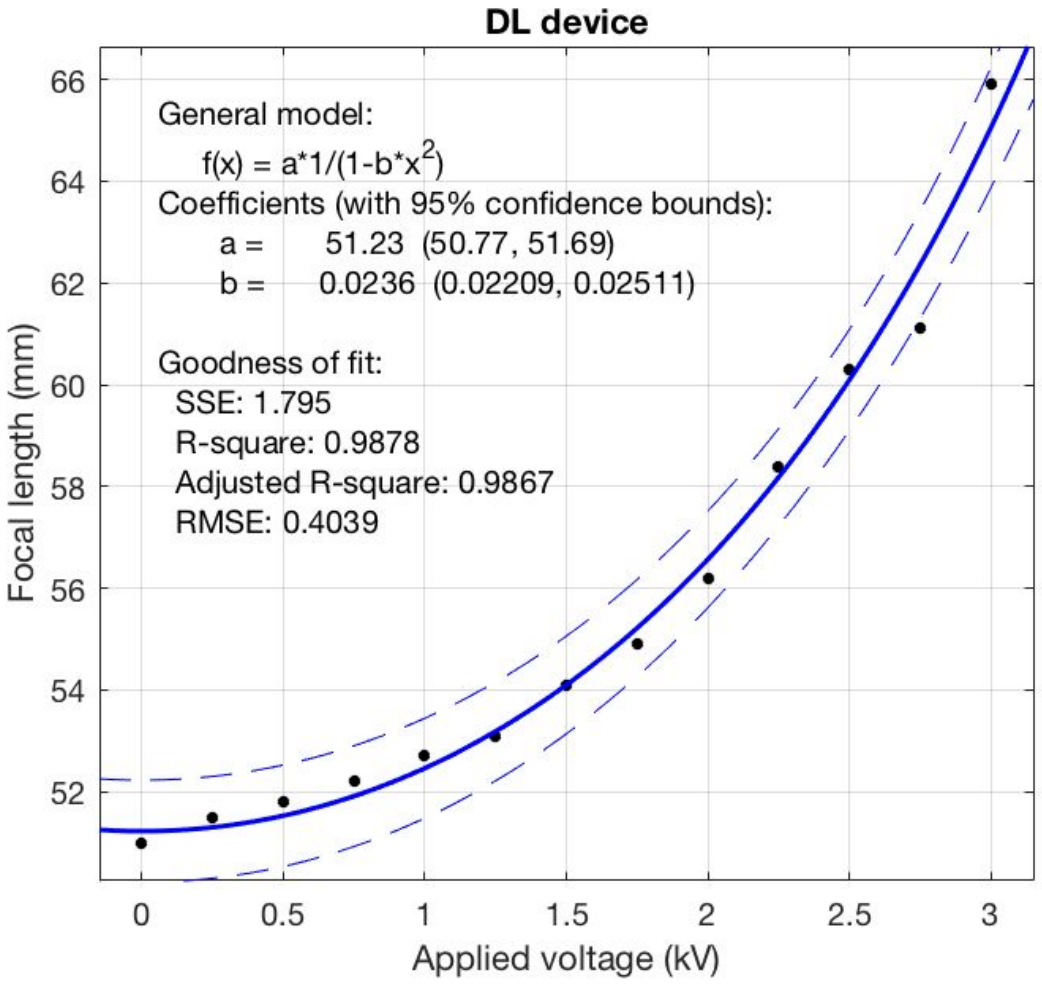

**Figure S7: Response time measurement.**

Measured rise and fall times for DL **(a-d)** and SL **(e-f)** devices. The measured data are plotted as black dots, the fits to exponential curves are shown as blue solid lines, and 95% confidence interval of the fit are shown as blue dashed lines. DL was actuated with a step voltage between 0 and 2 kV showing a **(a)** rise time of 33±3 ms and **(b)** fall time of 271±3 ms. DL was actuated with a step voltage between 0 and 3 kV showing a **(c)** rise time of 182±15 ms and **(d)** fall time of 105±13 ms. SL device actuated with a step voltage between 0 and 2 kV showing a **(e)** rise time of 327±7 ms and **(f)** fall time of 320±8 ms.

**a**

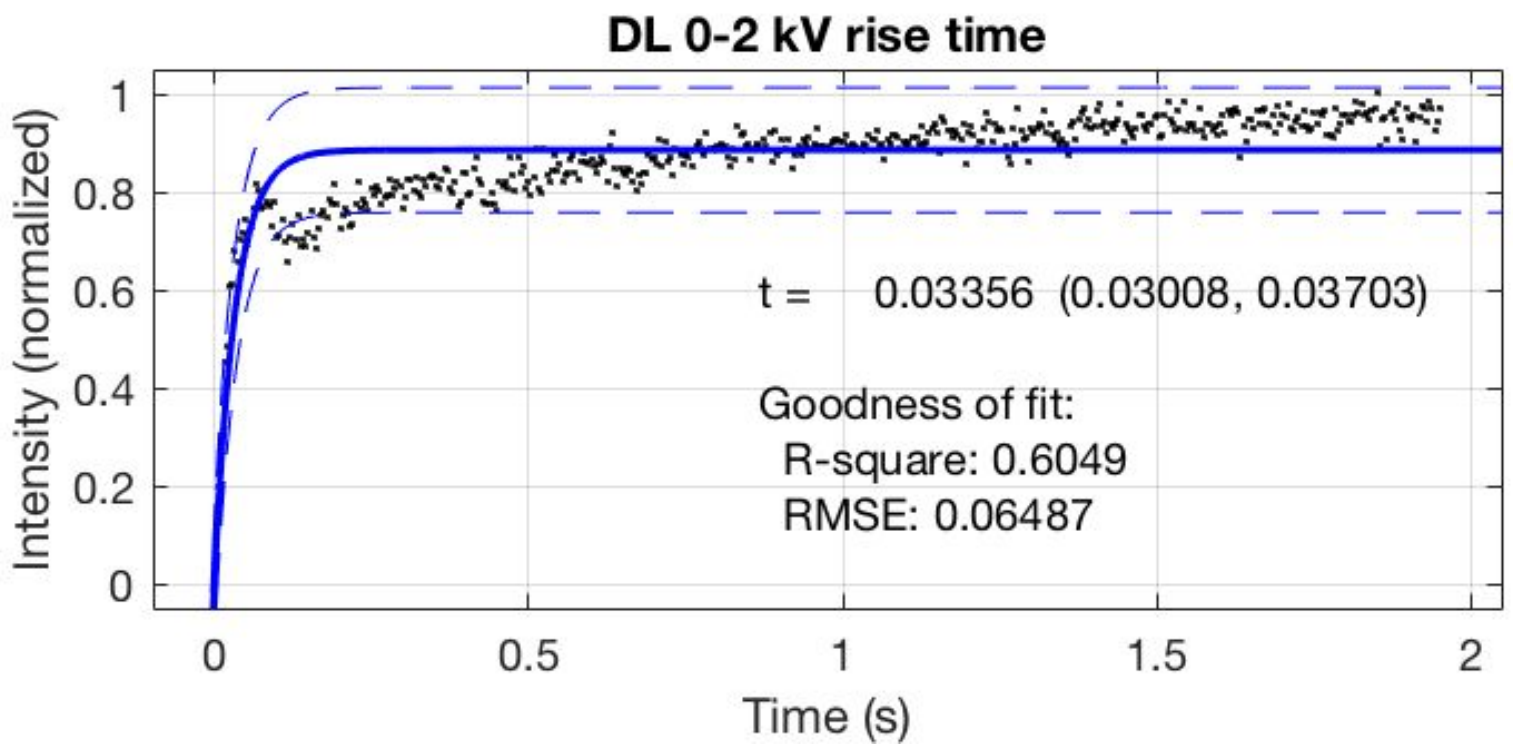

**b**

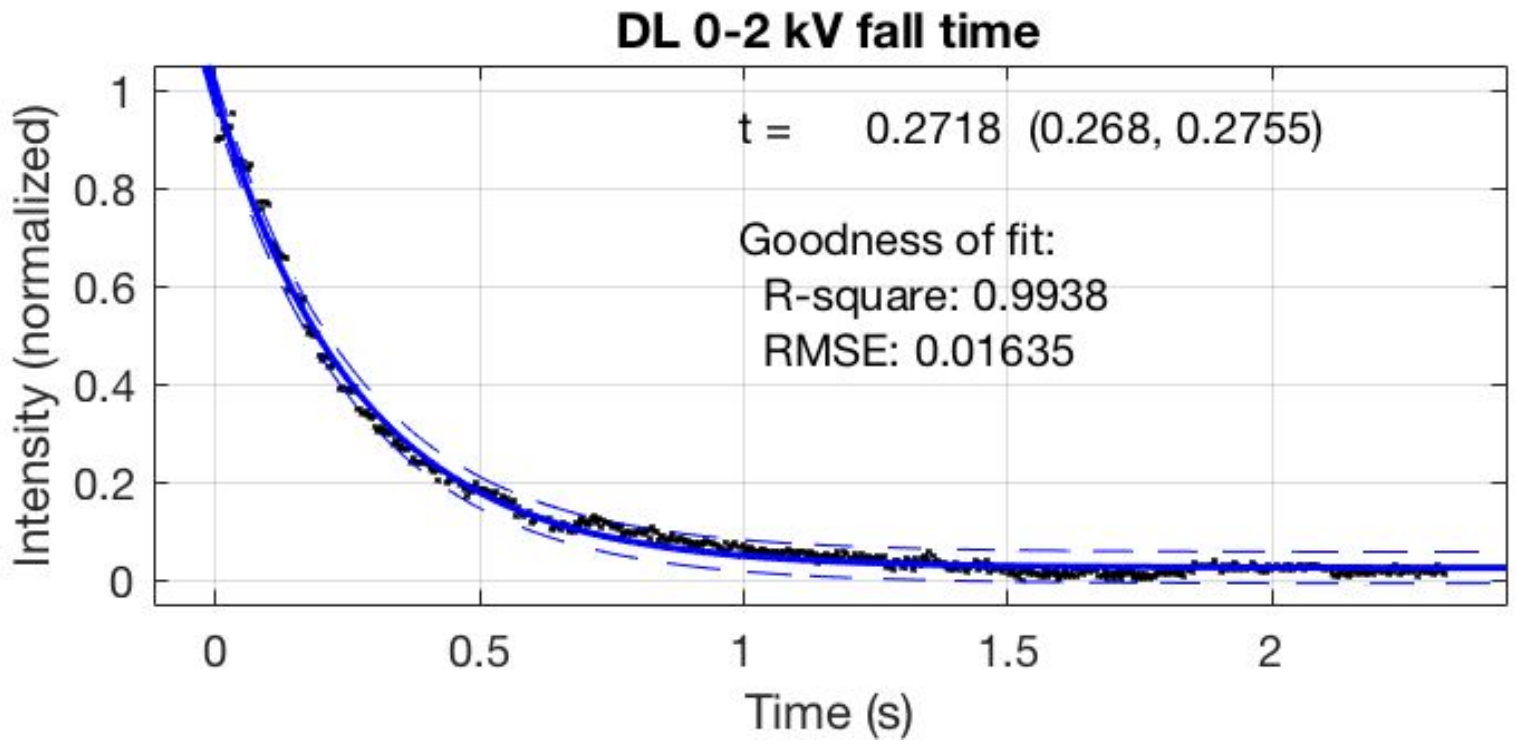

**c**

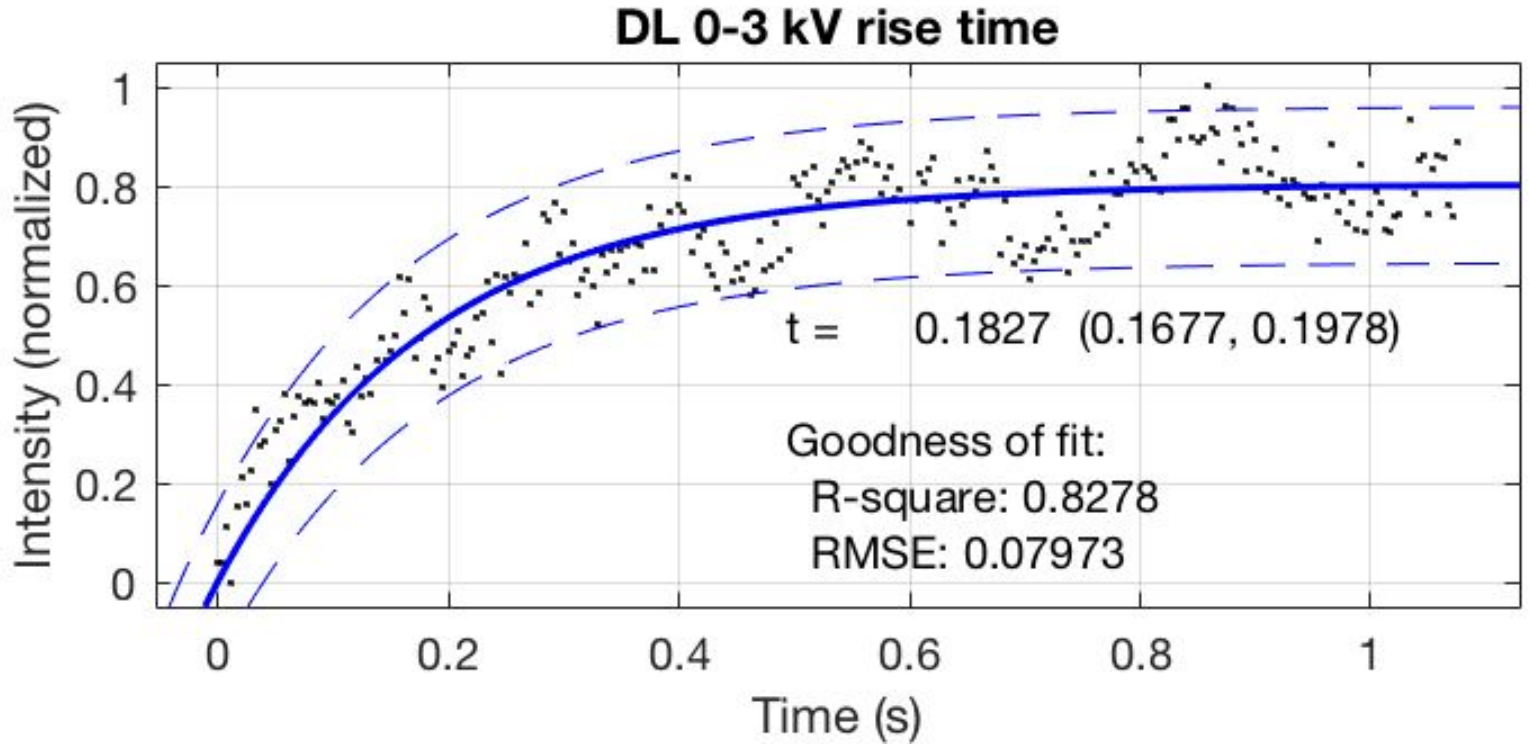

d

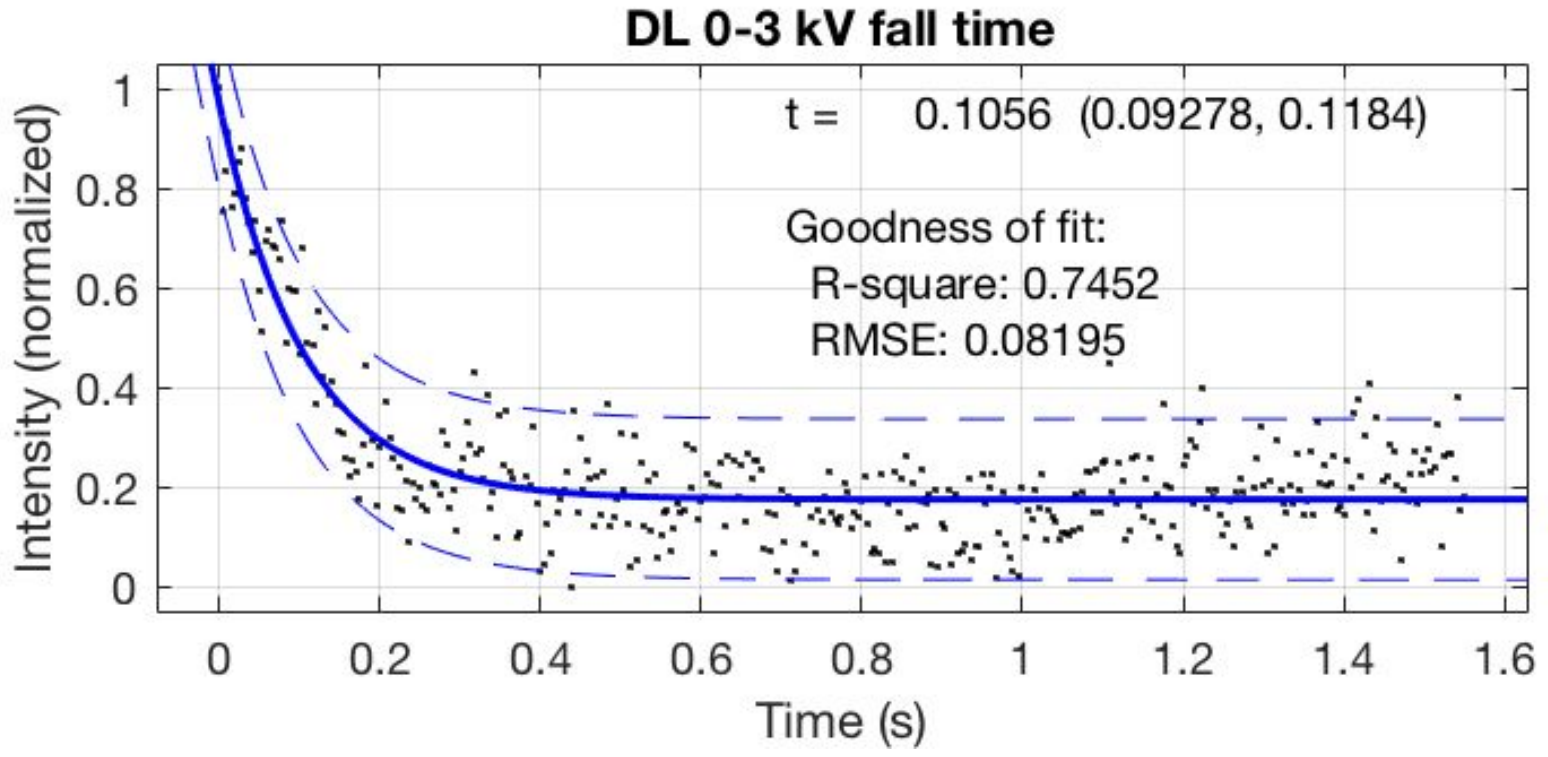

e

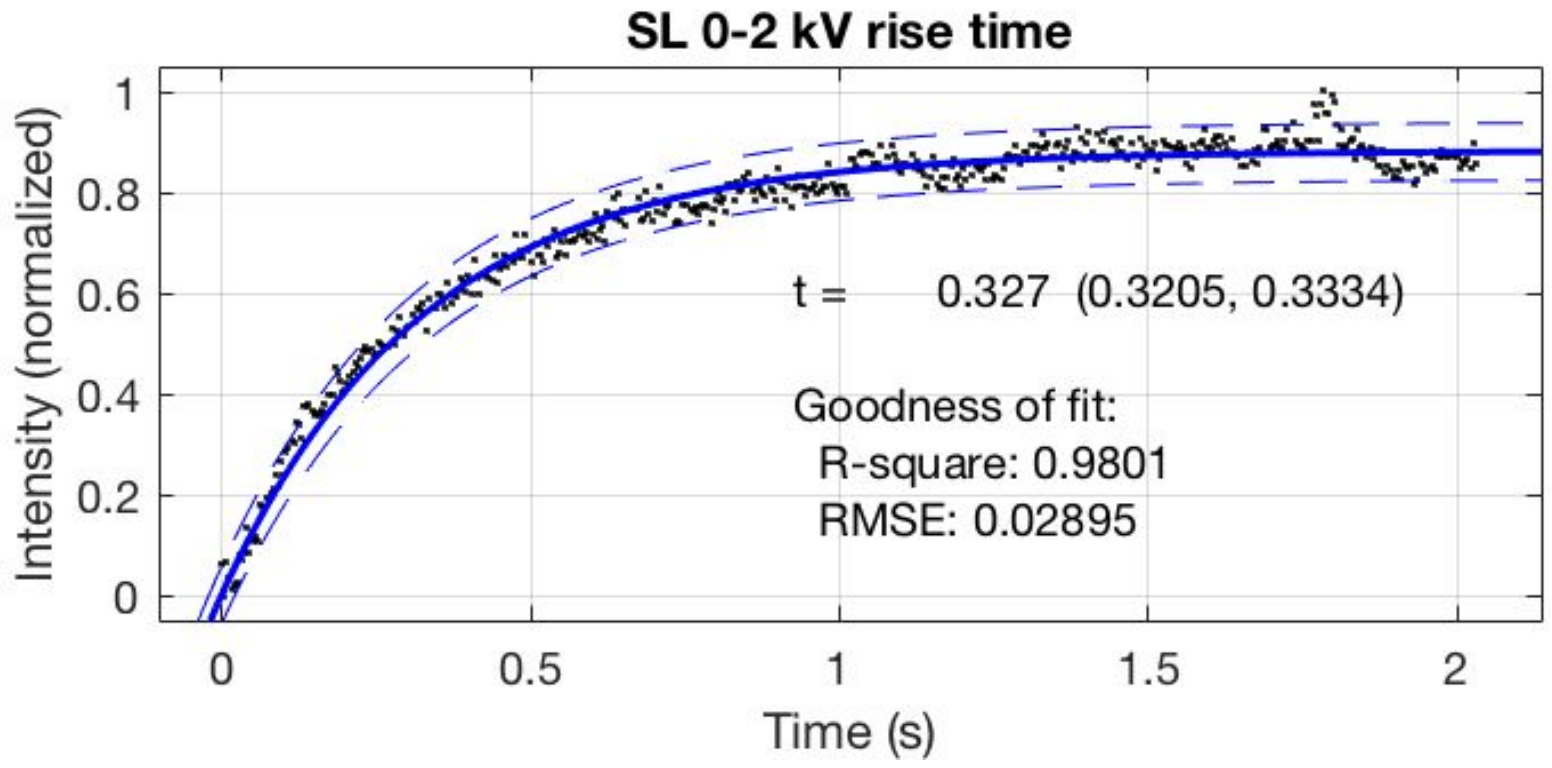

f

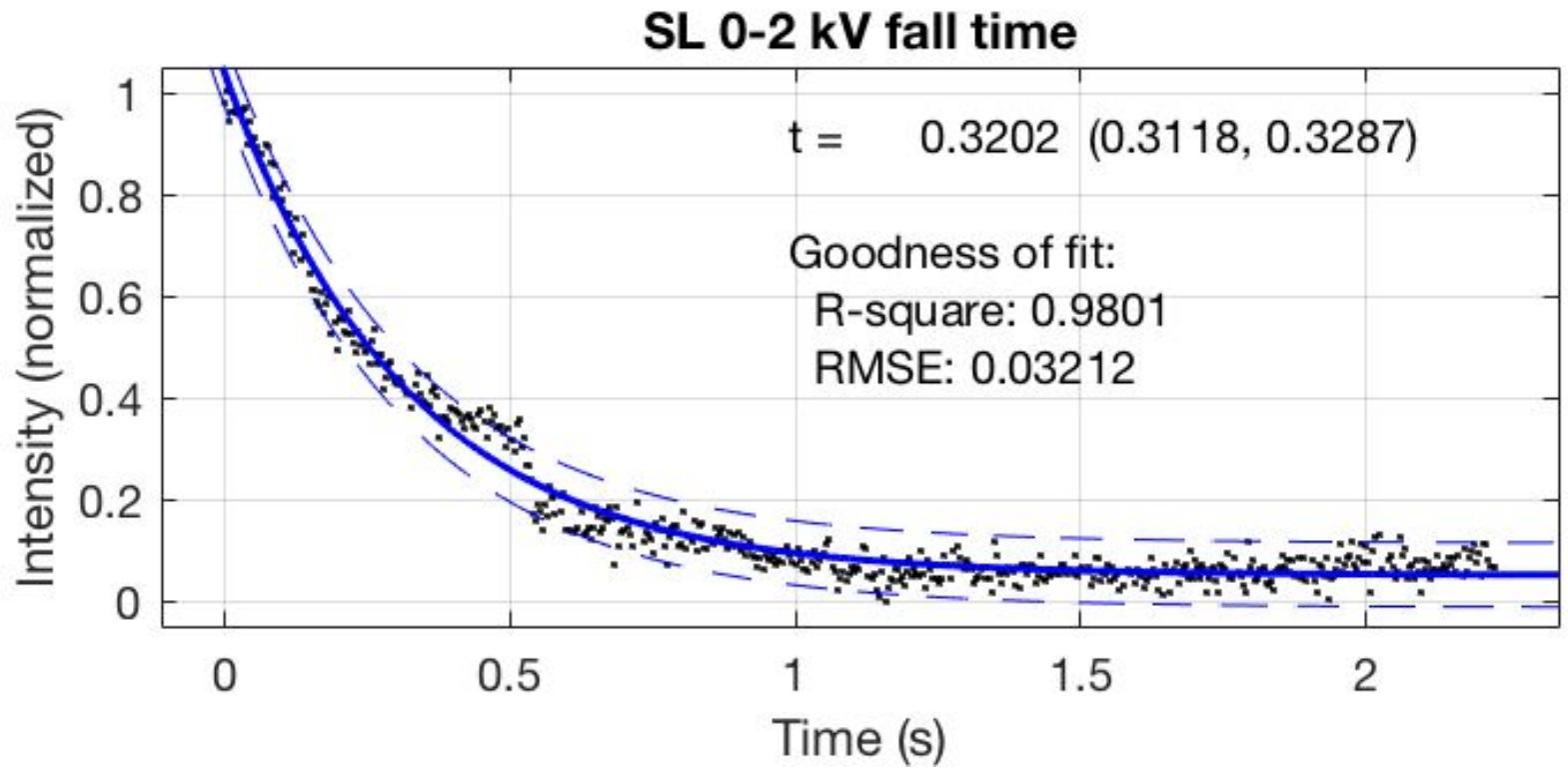

**Table S1: Wavefront shaping with the Zernike polynomials.**

The first 15 lowest order terms of the Zernike polynomial are presented. A series of Zernike terms can express any wavefront function or metasurface phase profile over a unit disk: $\phi(r,\theta) = \sum_j a_j Z_j(r,\theta)$, where $j$ is the Noll index and $a_j$ is the coefficient amplitude of $Z_j$. Also shown are the Zernike terms after uniform stretching, in which the radial coordinate transforms as $r \rightarrow r/s$. The stretched Zernike term is indicated as "re-scalable" if the $s$ dependence can be factored out cleanly (while ignoring any constant offset), effectively yielding a tunable coefficient of $Z_j$, in which the form of $Z_j$ is preserved when $s$ is varied, thus indicating that the particular term is a good candidate to be tuned via uniform stretching.

| Noll index (j) | Classical name | Radial degree (n) | Azimuthal degree (m) | Zernike term ($Z_j$ or $Z_n^m$) | Uniformly Stretched Zernike Term ($Z_j' = Z_j(r \rightarrow r/s)$) | Re-scalable? (Y/N) |
|---|---|---|---|---|---|---|
| 1 | Piston | 0 | 0 | 1 | 1 | N/A |
| 2 | Tip (X-Tilt) | 1 | 1 | $2r\cos(\theta)$ | $2r\cos(\theta)/s$ | Y |
| 3 | Tilt (Y-Tilt) | 1 | -1 | $2r\sin(\theta)$ | $2r\sin(\theta)/s$ | Y |
| 4 | Defocus | 2 | 0 | $\sqrt{3}(2r^2-1)$ | $\sqrt{3}(2r^2/s^2-1)$ | Y |
| 5 | Oblique astigmatism | 2 | -2 | $2\sqrt{6}r^2\sin(2\theta)$ | $2\sqrt{6}r^2\sin(2\theta)/s^2$ | Y |
| 6 | Vertical astigmatism | 2 | 2 | $2\sqrt{6}r^2\cos(2\theta)$ | $2\sqrt{6}r^2\cos(2\theta)/s^2$ | Y |
| 7 | Vertical coma | 3 | -1 | $\sqrt{8}(3r^3-2r)\sin\theta$ | $\sqrt{8}(3r^3/s^3-2r/s) sin\theta$ | N |
| 8 | Horizontal coma | 3 | 1 | $\sqrt{8}(3r^3-2r)\cos\theta$ | $\sqrt{8}(3r^3/s^3-2r/s) cos\theta$ | N |
| 9 | Vertical trefoil | 3 | -3 | $\sqrt{8}r^3\sin(3\theta)$ | $\sqrt{8}r^3\sin(3\theta)/s^3$ | Y |
| 10 | Oblique trefoil | 3 | 3 | $\sqrt{8}r^3\cos(3\theta)$ | $\sqrt{8}r^3\cos(3\theta)/s^3$ | Y |
| 11 | Primary spherical | 4 | 0 | $\sqrt{5}(6r^4-6r^2+1)$ | $\sqrt{5}(6r^4/s^4-6r^2/s^2+1)$ | N |
| 12 | Vertical secondary astigmatism | 4 | 2 | $\sqrt{10}(4r^4-3r^2)\cos(2\theta)$ | $\sqrt{10}(4r^4/s^4-3r^2/s^2)\cos(2\theta)$ | N |
| 13 | Oblique secondary astigmatism | 4 | -2 | $\sqrt{10}(4r^4-3r^2)\sin(2\theta)$ | $\sqrt{10}(4r^4/s^4-3r^2/s^2)\sin(2\theta)$ | N |
| 14 | Vertical quadrafoil | 4 | 4 | $2\sqrt{10}r^4\cos(4\theta)$ | $2\sqrt{10}r^4\cos(4\theta)/s^4$ | Y |
| 15 | Oblique quadrafoil | 4 | -4 | $2\sqrt{10}r^4\sin(4\theta)$ | $2\sqrt{10}r^4\sin(4\theta)/s^4$ | Y |

**Movie S1: Reliability test.**

A sinusoidal applied voltage (2 Hz) induces an oscillating focal spot size (via focal length change). The video was acquired at fixed distance from the lens.